# Automatic autism spectrum disorder detection using artificial intelligence methods with MRI neuroimaging: A review

**Parisa Moridian[1], Navid Ghassemi[2], Mahboobeh Jafari[3], Salam Salloum-Asfar[4], Delaram Sadeghi[5], Marjane Khodatars[5], Afshin Shoeibi[6,*], Abbas Khosravi[7], Sai Ho Ling[8], Abdulhamit Subasi[9,10], Roohallah Alizadehsani[7], Juan M. Gorriz[6], Sara A Abdulla[4,*], U. Rajendra Acharya[11,12,13]**

[1] Faculty of Engineering, Science and Research Branch, Islamic Azad University, Tehran, Iran.
[2] Computer Engineering Department, Ferdowsi University of Mashhad, Mashhad, Iran.
[3] Electrical and Computer Engineering Faculty, Semnan University, Semnan, Iran.
[4] Neurological Disorders Research Center, Qatar Biomedical Research Institute, Hamad Bin Khalifa University, Qatar Foundation, Doha, Qatar.
[5] Dept. of Medical Engineering, Mashhad Branch, Islamic Azad University, Mashhad, Iran.
[6] Data Science and Computational Intelligence Institute, University of Granada, Granada, Spain.
[7] Institute for Intelligent Systems Research and Innovations (IISRI), Deakin University, Geelong, VIC, Australia.
[8] Faculty of Engineering and IT, University of Technology Sydney (UTS), Ultimo, NSW, Australia.
[9] Institute of Biomedicine, Faculty of Medicine, University of Turku, Turku, Finland.
[10] Dept. of Computer Science, College of Engineering, Effat University, Jeddah, Saudi Arabia.
[11] Ngee Ann Polytechnic, Singapore, Singapore.
[12] Department of Biomedical Informatics and Medical Engineering, Asia University, Taichung, Taiwan.
[13] Department of Biomedical Engineering, School of Science and Technology, Singapore University of Social Sciences, Singapore.

* Correspondence: Afshin Shoeibi (Afshin.shoeibi@gmail.com) and Sara A Abdulla (saabdulla@hbku.edu.qa)

**Abstract:** Autism spectrum disorder (ASD) is a brain condition characterized by diverse signs and symptoms that appear in early childhood. ASD is also associated with communication deficits and repetitive behavior in affected individuals. Various ASD detection methods have been developed, including neuroimaging modalities and psychological tests. Among these methods, magnetic resonance imaging (MRI) imaging modalities are of paramount importance to physicians. Clinicians rely on MRI modalities to diagnose ASD accurately. The MRI modalities are non-invasive methods that include functional (fMRI) and structural (sMRI) neuroimaging methods. However, diagnosing ASD with fMRI and sMRI for specialists is often laborious and time-consuming; therefore, several computer-aided design systems (CADS) based on artificial intelligence (AI) have been developed to assist specialist physicians. Conventional machine learning (ML) and deep learning (DL) are the most popular schemes of AI used for diagnosing ASD. This study aims to review the automated detection of ASD using AI. We review several CADS that have been developed using ML techniques for the automated diagnosis of ASD using MRI modalities. There has been very limited work on the use of DL techniques to develop automated diagnostic models for ASD. A summary of the studies developed using DL is provided in the Supplementary Appendix. Then, the challenges encountered during the automated diagnosis of ASD using MRI and AI techniques are described in detail. Additionally, a graphical comparison of studies using ML and DL to diagnose ASD automatically is discussed. We suggest future approaches to detecting ASDs using AI techniques and MRI neuroimaging.

# 1. Introduction

A complex intricate network of millions of neurons is responsible for monitoring and controlling each part of the human body and brain [1-3]. These networks consist of many neurons that need to be directly interconnected and synchronized [4-5]. It has been suggested that certain disorders in the human body arise when brain networks are incorrectly connected to manage a specific activity [6-9]. Disorders of this type can be classified into three groups based on their psychological or neural characteristics and threaten the health of many individuals across the globe. Autism spectrum disorder (ASD) [10], schizophrenia [11], attention deficit hyperactivity disorder (ADHD) [12], epilepsy [13], Parkinson's disease [14], and bipolar disorder (BD) [15] are some of the most known neurodevelopmental disorders.

ASD is a neurodevelopmental disorder that manifests in childhood and causes a variety of challenges to individuals [1]. Those with ASD have difficulties with verbal and non-verbal communication, cognitive skills, social behavior, and entertaining activities. ASD begins in the early stages of embryonic development, according to research results. Autism is thought to be caused by specific signal patterns in the cortex, abnormalities in the immune system, growth hormone fluctuations, and abnormalities in the neural mirror system in the embryonic stage [16-17]. The overall ASD prevalence is one in 44 children aged 8 years, and ASD is around 4 times as prevalent among boys as among girls. [18-19]. In addition to lifelong social and adaptive disorders, one of the major consequences of autism is its negative impact on quality of life. Therefore, early diagnosis and treatment of ASD are of paramount importance for improving the quality of life of ASD children and their families [20].

According to the DSM-3, mental health professionals originally divided autism into five categories, including Asperger's syndrome, Rett syndrome, childhood disintegrative disorder (CDD), autistic disorder, and Pervasive developmental disorder-not otherwise specified (PDD-NOS) [21-22]. Using this method, physicians observed the symptoms of autistic individuals and compared their observations to those in the DSM-3 to diagnose the specific type of autism [21-23]. In 2013, the DSM-5 was published, making significant changes to the categorization of autism [24]. DSM-5 categorizes autism severity into three levels, and more information is given in [24]. According to DSM-5, the lower the severity level of autism, the less support the child requires. Autism individuals with the second and third severity levels show moderate to severe symptoms and therefore require more frequent support. Although the DSM-5 provides explanations of the autism spectrum, these explanations are incomplete and do not provide guidance on the specific support that may be required by autistic children. In addition, some individuals simply do not fall into any of these categories and ASD can change and intensify over time [24-25].

Neuroimaging techniques are one group of methods used for diagnosing neurological and mental disorders such as ASD. These methods comprise structural and functional neuroimaging modalities, which are of special interest to physicians, particularly in the diagnosis of various brain disorders [26-27]. The fMRI is one of the major functional neuroimaging methods that records data in a non-invasive manner. fMRI has a high spatial resolution, making it an excellent method for examining functional connectivity in the brain. fMRI data is classified into two categories: T-fMRI and rs-fMRI. Furthermore, fMRI data are composed of a 4-dimensional tensor, which permits the 3D volume of the brain to be segmented into smaller areas, and the activity of each area is recorded for a predetermined time period. Although fMRI has provided satisfactory results in the diagnosis of a variety of brain disorders, these techniques are costly and too sensitive to motion artifacts [29] [34].

sMRI and DTI have been used to examine brain anatomy and the interaction between brain regions, respectively. The structural neuroimaging modalities offer the advantage of cost-effectiveness and the availability of imaging protocols in most treatment facilities [34]. Physicians use sMRI modalities to diagnose autism in autistic individuals using i) geometric features and ii) volumetric features, which physicians have used to demonstrate that autistic people demonstrate superior brain development in comparison to normal people [30-33]. Hazlett et al. [35] studied the brain structure of 51 autistic children and 25 normal children (1.5-3 years of age). Their findings indicated that the Cerebellum white matter volume of autistic children was 2-4 times greater than that of normal children.

Although MRIs offer many advantages, MRI artifacts reduce the accuracy with which clinicians are able to diagnose autism. Additionally, ASD individuals' MRI data is recorded with multiple slices and different protocols. Consequently, it takes a considerable amount of time to examine all MRI slices accurately, and clinicians should be highly precise. The fatigue of the physician may lead to an incorrect diagnosis of ASD in many cases. In addition, MRI data is problematic because most physicians are inexperienced in interpreting these images and may find it difficult to diagnose ASD in its early stages.

In order to improve the accuracy of ASD diagnosis, AI techniques can be used. The use of AI in diagnosing various diseases has been studied [36-38]. Several studies have demonstrated that AI techniques, along with MRI neuroimaging modalities, increase the accuracy of ASD diagnosis [36-37]. An increasing number of studies have been conducted on the detection of ASD using ML and DL methods. Researchers first demonstrated that ASD can be diagnosed from ML using MRI neuroimaging technologies [38]. Feature extraction, dimension reduction, and classification algorithms in CADS based on ML algorithms are selected through trial and error [39-40]. Choosing an appropriate algorithm for each CADS section can be challenging without adequate knowledge of AI [39-41]. Furthermore, ML techniques are not suitable for small data sets [42]. These methods therefore do not contribute to the development of software for the detection of ASDs using MRI neuroimaging modalities.

A variety of studies are being conducted in order to diagnose various diseases and disorders by using these methods in order to overcome the challenges inherent in ML techniques [43-46]. In contrast to ML methods, DL uses deep layers for feature extraction and classification and require fewer implementation steps in the diagnosis of ASD [47]. Furthermore, DL-based CADS can be more efficient and accurate with large input data.

An overview of studies relating to the detection of ASD using MRI neuroimaging methods is presented in this comprehensive systematic review. The first step was to review systematically all publications on ASD detection using MRI modalities and ML techniques. A recent review by the authors of this review discussed the use of different neuroimaging modalities and DL architectures to detect ASD [6]. Appendix A presents a review paper describing ASD detection in different neuroimaging modalities using DL techniques to make a comparison between ML and DL studies.

The following sections describe the following. Section 2 is search Strategy based on PRISMA guideline. In section 3, the review papers in AI techniques for ASD diagnosis are reviewed. Section 4 describes the CADS based on AI to detect ASD from MRI neuroimaging images. In section 5, a comparison between ML and DL studies to ASD detection using MRI modalities is presented. Section 6 examines the most critical challenges for detecting ASD using AI methods. Future directions and conclusion sections are presented in sections 7 and 8, respectively.

## 2. Search Strategy Based on PRISMA Guideline

The PRISMA protocol was used to select and review papers in this study [11]. Papers on the diagnosis of ASD by MRI modalities and AI models (ML and DL) published from 2016 to 2022 were included in this study. In this review paper, various citation databases, including IEEE, Wiley, Frontiers, ScienceDirect, SpringerLink, ACM, and ArXiv were used to search for papers in the field of ASD detection. Furthermore,

Google Scholar has been used to search for the article entirety. Here are the keywords, including "ASD classification," "Feature extraction", "fMRI". "sMRI" and "Autism Spectrum Disorder," which were used to search for articles relating to the diagnosis of ASD using ML algorithms. To search for articles related to DL, the keywords used were "Autism Spectrum Disorder", "ASD", fMRI", "sMRI", and "Deep Learning".

As stated above, papers were selected and reviewed based on the PRISMA protocol at three different levels. In the first level, 34 out of 316 downloaded papers were eliminated as they were out of the scope of this study. Then, 28 papers were also excluded as they did not use MRI datasets in the ASD diagnosis, followed by excluding further 21 papers due to no use of AI techniques. Therefore, 233 papers were finally selected and used in this review paper. Figure (1) shows the selection procedure of papers based on the PRISMA protocol on three levels. The key criteria for the inclusion and exclusion of papers on the ASD diagnosis based on the PRISMA protocol are shown in Table (1).

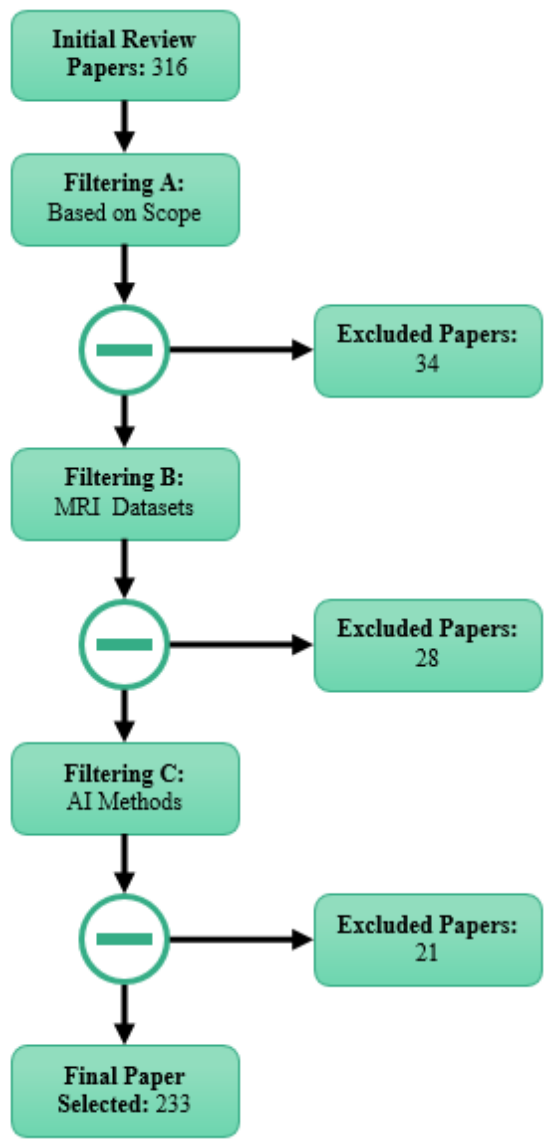

Fig 1. Papers selection process based on the PRISMA guidelines.

## 3. Artificial Intelligence Techniques for ASD diagnosis

For researchers in cognitive sciences, autism is a well-recognized neurodevelopmental disorder with a high prevalence in recent years. Challenges in the ASD diagnosis for physicians have resulted in extensive studies on this brain disorder. Scholars in AI, and cognitive sciences seek to develop a real diagnostic tool for ASD using various AI techniques. Accordingly, extensive studies have focused on ASD diagnosis using neuroimaging modalities and AI techniques, outlined in this section by reviewing articles in the field of ASD diagnosis using these techniques.

Table 1. The exclusion and inclusion criteria for diagnosis of ASD.

| Inclusion | Exclusion |
| --- | --- |
| 1. sMRI neuroimaging modalities | 1. Treatment of ASD |
| 2. fMRI neuroimaging modalities | 2. Clinical methods for ASD treatment |
| 3. Different Types of Autism | 3. Rehabilitation systems for ASD detection (Without AI techniques) |
| 4. DL models | |
| 5. Feature extraction methods | |
| 6. Dimension reduction methods | |
| 7. Classification methods | |

Pagnozzi et al. [344] reviewed 123 articles on ASD diagnosis using sMRI modalities and reported further developments in some brain areas of ASD individuals than those of HC. They also explained that ASD caused changes in the structure of patients' brains, including increased volume of frontal and temporal lobes, increased thickness of the frontal cortex, and increased volume of cerebrospinal fluid. This study assists scholars in applying AI techniques in ASD diagnosis from sMRI modalities in future studies.

Nogay et al. [345] published a review article on ASD diagnosis using brain imaging and ML techniques. They reviewed studies on ASD diagnosis for sMRI, fMRI, and combined data using ML techniques and found a higher accuracy of ASD diagnosis at younger ages. They hope to develop a practical ASD diagnostic tool based on ML techniques from MRI modalities.

In another study, Xu et al. [38] reported methods and tools associated with ASD diagnosis from MRI data based on ML techniques. Initially, they introduced the most important ML-based algorithms, including feature extraction, feature selection and reduction, training and test models, and, finally, evaluation parameters.

Parlett-Pellerit et al. [37] reviewed studies on unsupervised ML techniques for ASD diagnosis. In this study, various clinical data and medical imaging data were discussed for ASD diagnosis using unsupervised ML techniques.

The most important feature selection and classification algorithms for ASD diagnosis were studied in Rahman et al. [346] paper. Their input data comprises various psychological tests such as ADOS and MRI modalities. They claimed that this study could assist scholars in developing future studies on ADS diagnosis.

A review article on the diagnosis of ASD and ADHD using AI techniques was published by Eslami et al. [347]. They discussed DL and ML-based studies on ASD and ADHD diagnosis from MRI modalities and the most important AI techniques (DL and ML). In the ML section, the authors presented the most important feature extraction techniques, such as dynamic effective connectivity, and outlined various popular DL techniques.

Khodatars et al. [6] presented a review paper on ASD diagnosis and rehabilitation using DL techniques. They initially introduced the public neuroimaging modalities datasets, such as MRI, pre-processing techniques, and DL models, an ASD diagnosis. Then, they summarized the studies conducted in this field in a table. They also discussed studies in the field of autism rehabilitation using DL techniques.

In this section, the most important review papers on ASD diagnosis from various data and AI techniques were discussed. In our study, ASD diagnosis papers using MRI data and various AI techniques (ML and DL) were reviewed. This paper reports all ASD diagnosis articles from 2010 to 2022. Also, the most important challenges and future works for diagnosing ASD from MRI modalities are presented. To the best of our knowledge, no similar review article has been provided so far, and our review article has outstanding innovations.

## 4. CADS for ADS diagnosis by MRI Neuroimaging Modalities

The application of CADS based on AI techniques is presented in this section, and it is illustrated in Figure (2). The steps involved in CADS using ML for ASD detection are outlined in Figure (2). As shown in figure (2), CADS input consists of datasets containing MRI modalities. Standard preprocessing (low-level) methods for MRI neuroimaging modalities were demonstrated as a second step. Next, we will discuss these preprocessing methods for MRI neuroimaging modalities. The third step involves feature extraction. Feature reduction or selection techniques (dimension reduction) are considered to be the fourth step in the CADS based on ML. The final step involves the use of classification algorithms. The only difference between DL-based and ML-based CADS is the feature extraction to the classification step. This procedure is carried out in deep layers in CADS based on DL. This enables the extraction of features from MRI data without the user's intervention. Moreover, in CADS based on DL, diagnostics of ASD may be possible in case there are more input data, allowing the development of actual software for the detection of ASD. The details of ASD detection from MRI neuroimaging modalities using DL architectures are given in Appendix (A). Here we present the details of CADS based on ML, along with some of the most important algorithms in each section.

### 4.1. MRI Neuroimaging ASD Datasets

Various MRI modalities datasets for ASD diagnosis are available to researchers, including UCI [348], NDAR [349], AGRE [350], NRGR [351], GEO [352], SSC [353], Simons VIP [354], and autism brain imaging data exchange (ABIDE) [6]. Tables (2) and (3) summarize studies of ASD diagnosis using ML and DL techniques. As can be seen, the ABIDE database has a special place in research. ABIDE is recognized as the most complete and freely available database of MRI modalities for the automatic diagnosis of ASD [6]. This dataset has two parts, ABIDE 1 and ABIDE-II, containing sMRI data, rs-fMRI data, and phenotypic data. 1112 datasets are involved in ABIDE I, and 1114 datasets are included in ABIDE II [6]. ABIDE 1 also contains preprocessed data from MRI modalities for research, known as the preprocessed connectomes project (PCP) [6]. Additionally, other available datasets, such as NDAR, UCI, and NRGR, have been used in ASD diagnostic. The results show that these datasets have been able to achieve satisfactory results. The datasets used for each study are summarized in Tables (2) and (3).

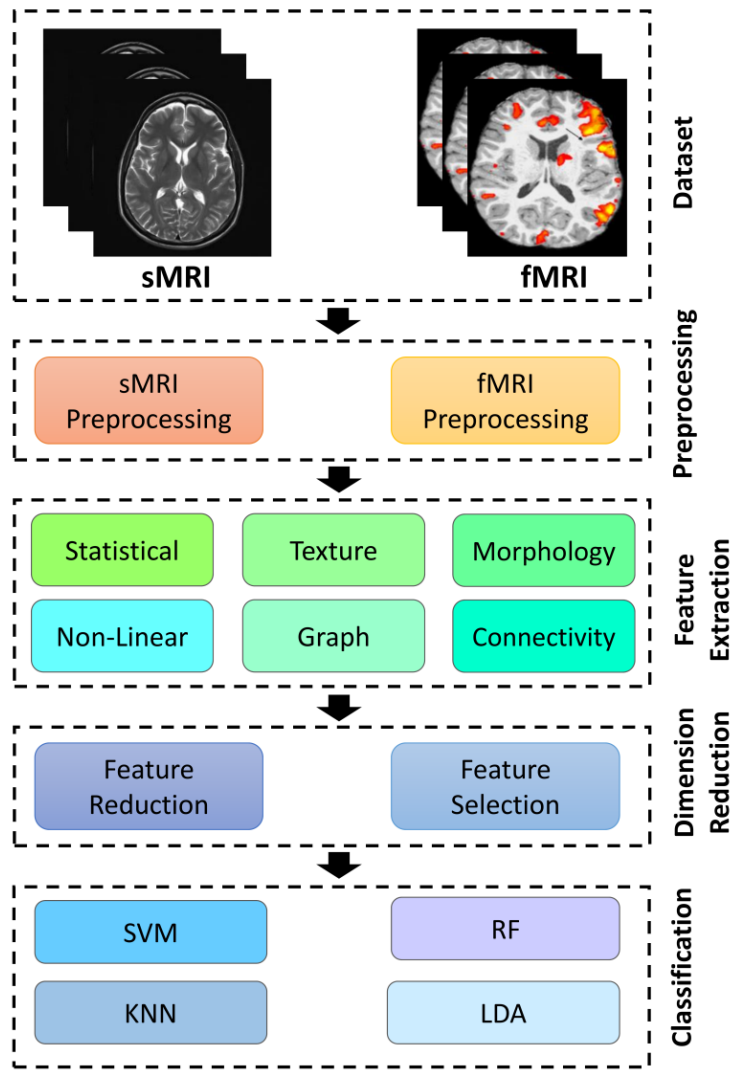

Fig. 2. Block diagram of CADS- based on ML techniques for automated ASD diagnosis.

### 4.2. Preprocessing Techniques for fMRI and sMRI Modalities

Preprocessing techniques are needed to help CADS to achieve promising results. The sMRI and fMRI neuroimaging modalities have implemented fixed preprocessing steps using different software packages. The most common software packages are brain extraction tools (BET) [48], FMRIB software libraries (FSL), statistical parametric mapping (SPM), and FreeSurfer [6]. In the following, the standard preprocessing steps for fMRI and sMRI neuroimaging modalities are presented. Some of them are common for both fMRI and sMRI modalities, so we will introduce them in the fMRI-related section. Figure (3) shows the standard fMRI and sMRI techniques. Also, the pipeline preprocessing techniques for ABIDE datasets will be introduced in another section.

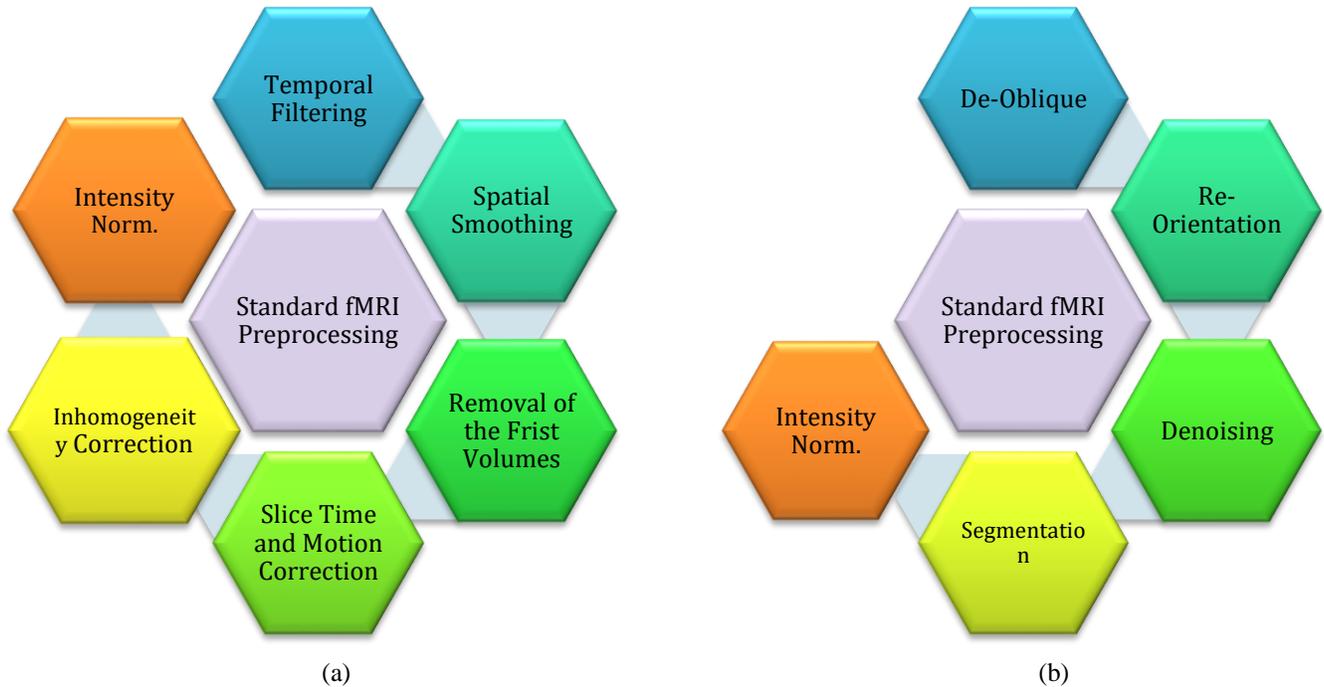

(a)          (b)

Fig. 3. Standard preprocessing methods for MRI neuroimaging modalities: a) Preprocessing for fMRI data, b) preprocessing for sMRI data.

The standard Preprocessing is a necessary step in fMRI, and if preprocessing is not carried out properly, it will lead to reduced performance of automated diagnosis of ASD. This step aims to extract regions suspected of having ASD and examine the function of brain neurons in those regions. The preprocessing steps of fMRI include delineation of the brain region, removal of the first few volumes, slice timing correction, inhomogeneity correction, motion correction, intensity normalization, temporal filtering, spatial smoothing, and ultimately registration standard atlas [6]. As mentioned earlier, this step is usually carried out by a toolbox, including BET [6], FSL [6], SPM [6][50], FreeSurfer [6][51], etc. In reference [6], the details for standard preprocessing steps of fMRI modalities are elaborately explained.

The preprocessing of sMRI data extracts helps physicians examine regions with suspected ASD more accurately. Besides, low-level sMRI preprocessing methods help AI-based CADS to process important information. This process increases the accuracy and efficiency of diagnosis of ASD CADS. The most important standard sMRI covers intensity standardization, de-oblique, re-orientation, Denoising, and segmentation [6]. In reference [6], each step of standard preprocessing for sMRI modalities is explained.

### 4.2.1. Pipeline Methods

The pipelines are a preprocessed version of ABIDE data using standard preprocessing procedures, which researchers can use to avoid the problems of variations in the output between different studies as a result of preprocessing. The most popular ABIDE pipelines include neuroimaging analysis kit (NIAK), data processing assistant for rs- fMRI (DPARSF), the configurable pipeline for the analysis of connectomes (CPAC), and connectome computation system (CCS) [6].

### 4.3. Feature Extraction

Representing data in a manner that allows ML algorithms to reason about them is the key to any related research. If this is not done, high performance cannot be achieved. Most commonly used feature extraction schemes for fMRI and sMRI are statistical, texture, morphological, non-linear, graph, functional connectivity, and structural connectivity schemes.

- **Statistical Features**

ASD is typically detected with MRI modalities using statistical features, the most straightforward group of features. Despite their simplicity, these features are usually informative and can also serve as a benchmark for evaluating other methods of feature extraction as well. Additionally, the process of extracting these features is not time-consuming in comparison to other methods. However, these methods do not reveal non-linear or subtle patterns in data. Using statistical features for ASD diagnosis, Dekhil et al. [80] extracted various statistical features from MRI data and then applied KNN and SVM algorithms in the classification step. The authors reported 81% accuracy.

- **Texture Features**

As a group of features, spatial patterns form an indispensable group, possibly the most important group, since the cognitive system of the human is mostly dependent on them. Gray-level co-occurrence matrix (GLCM) [49] feature extraction is one of the most widely used methods in various research studies [94] among various textures-based features. Haweel et al. [67] presented an ASD diagnostic method based on MRI data. Texture features and the RFE technique were used in the feature extraction and feature selection steps. Then, the authors used the RF technique for classifying features and reached an accuracy of 72%. In another study, scholars used various methods, such as Haralick, in the feature extraction step from sMRI data. Then, the authors tested different feature selection methods and reached an accuracy of 72.5%.

- **Morphological Features**

Morphological operation is an important feature extraction technique frequently used in image processing [366]. In these methods, features are extracted from the appearance and shape of the image. Morphological operation is often used to process binary images, but they can also be used for gray and color-level images [367]. Morphological features are also commonly used for diagnosing brain diseases from sMRI modalities. Zheng et al. [75] proposed the idea of ASD diagnosis using morphological features. After feature extraction, RFE and SVM were tested for feature selection and classification, respectively. An accuracy of 78.63% was obtained

- **Non-Linear Features**

A non-linear characteristic of biological data is emphasized when considering non-linear features. The performance of CADS for ASD is significantly enhanced through the use of these features [50]. In reference [103], nonlinear-based features of likelihood are used to detect autism using MRI neuroimaging methods. Entropies are one of the most important nonlinear methods that are widely used to extract features from signals and brain images [368]. Functional imaging modalities are nonlinear and chaotic, which has led researchers to use entropy-based nonlinear features to diagnose brain disorders [330] [369]. Zhang et al. [133] introduced a novel ASD diagnostic method using fMRI data and a new entropy method. This study initially used fast entropy for feature extraction from preprocessed fMRI data. Then, they used the SVM algorithm for feature classification and obtained satisfactory results.

- **Graph Features**

This group of features is highly relevant to the analysis of MRI data. Graph-based features are derived first by shaping the data into a graph, and then, from the constructed graph, local and global features are extracted [51]. Researchers have explored graph features to diagnose ASD using fMRI data in many studies.

Bi et al. [70] employed rs-fMRI from the ABIDE database for ASD diagnosis using graph and genetic-evolutionary random SVM cluster (GERSVMC) for feature extraction and classification steps, respectively, and obtained an accuracy of 62%. Saad et al. [72] presented an ASD diagnostic method based on graph features in another study. After feature extraction via the graph method, PCA and SVM techniques were used for feature reduction and classification, which resulted in an accuracy of 75% for ASD diagnosis.

- **Connectivity Matrix**

In order to process sMRI and fMRI neuroimaging images, feature extraction methods based on connectivity matrix methods are typically employed [52-53]. Such features provide information about the brain's structure and function. The functional connectivity matrix (FCM) [56-57] and structural connectivity matrix (SCM) [58-59] are the measures employed for fMRI and sMRI modalities, respectively. Connectivity features are mostly used in diagnosing brain disorders. Tables (2) and (3) summarize studies on ASD diagnosis from MRI modalities using various AI techniques. Table (2) shows that connectivity methods are most frequently used for feature extraction from MRI modalities. Liu et al. [84] used dynamic functional connectivity (DFC) in the feature extraction step of rs-fMRI data. The feature selection step was also conducted by the MTFS-EM method. Finally, they used the SVM method for classification and obtained an accuracy of 76.84%. In another study, Mathur et al. [125] utilized DFC and static functional connectivity (SFC) in the feature extraction step. Then, the SVM was tested for connectivity-based classification of features. Authors could finally obtain satisfactory results for ASD diagnosis using connectivity features.

## 4.4. Feature Reduction / Selection Methods

It has been shown that increasing the number of extracted features can help algorithms to represent data in a more meaningful and robust way; however, the curse of dimensionality [54] may cause it to backfire and reduce performance. Several methods for reducing dimensionality and selecting features have been proposed in order to prevent this from occurring. In addition, these methods are widely used to increase the performance of CADS for the detection of autism spectrum disorders. A number of methods have previously been used in research papers, including principal component analysis (PCA) [55], recursive feature elimination (RFE) [56], T-test [57], autoencoder (AE) [58], conditional random forest (CRF) [59], Chi-squared [60], and least absolute shrinkage and selection operator (LASSO) [61]. The following is a brief description of these methods.

- **PCA**

PCA is arguably the most common dimensionality reduction method [55]. It works by finding and representing data by the principal components, i.e., the vectors that preserve the most data variance. One of the benefits of PCA is its ability to find a minimal number of features required to preserve a specified variance ratio [55]. Principal component analysis (PCA) is one of the most popular feature reduction techniques in medical applications and has also been used innASD diagnosis research [72], [89], [110], [150].

- **RFE**

Recursive feature elimination is more of a wrapper-type algorithm, meaning that it is applied to a classification algorithm to find the best subset of features. As the name explains, this algorithm works by eliminating features one by one to reach the optimal number. First, a classification algorithm is trained on the dataset, ranking feature importances. The least important feature is then eliminated, and the process is repeated until the number of remaining features matches the desired number [56]. Haweel et al. [67] proposed a novel ASD diagnostic method using the GLM feature extraction technique. After feature extraction from MRI data, the RFE technique was used for feature reduction. The RF method was also tested in the classification step with an accuracy of 72%.

- **T-Test**

To find the best set of features, T-Test calculates a score for each feature, then ranks them based on that score and picks the top features as selected ones. The score shows whether the values of a feature for a class are significantly different from those for another class by calculating the mean and standard division (STD) of each feature in each class [57]. A new ASD diagnostic method from MRI data was introduced by Sartipi et al. [71]. First, the graph technique was used for feature extraction from sMRI modalities. Then, they applied the t-test and SVM algorithms for the feature selection and classification steps and acquired an accuracy of 75%.

- **Chi-Squared**

Chi-Square is suitable when the features are categorical and the target variable is also categorical, such as classification. Chi-Squared measures the degree of association between two variables; thus, features that have a connection with the targets can be picked [60]. When the features are numerical, we can use a T-Test, or Chi-Square can be used for the numerical variable by discretizing them [60]. In reference [80]. The authors proposed a new ASD diagnostic method using various ML techniques from MRI data. Various methods were used for feature extraction. Then, the Chi-squared method was tested for the feature selection step. Next, the LR classification algorithm was applied, which resulted in a promising performance.

- **LASSO**

Lasso is mainly a regression method; however, this algorithm can also be used for feature selection [61]. Notably, linear regression with L1 regularization is called Lasso. After training, the lasso assigns a weight to each feature for the regression [61]. Using these weights, there are two methods to pick the best features, first, pick the K highest valued weights; second, pick all the weights which have a value higher than a specified threshold [61].Fedro et al. [113] proposed a new ASD diagnostic method based on Hons and Lon features. Their paper used LASSO and SVM methods for feature selection and classification, respectively. They reported an accuracy of 81%.

**4.5. Classification Methods**

This section discusses the various classification algorithms used in CADS for ASD. As mentioned earlier, classification is the last step in a CADS based on ML methods. Support vector machine (SVM) [62-63], linear discriminant analysis (LDA) [65], k-nearest neighbor (KNN) [66], and random forest (RF) [64] are arguably among the most popular methods used in CADS created for ASD. Tables (2) and (3) show the classification algorithms used for ASD detection. A brief summary of classification algorithms used for automated detection of ASD is presented below.

- **SVM**

Support vector machines (SVMs) are among the oldest classification which has been widely used in many applications [62-63]. SVM tries to find the best hyperplane to separate data points; however, it only needs the dot product between every two data points [62-63]. Consequently, to transform data into another space, only a function that gives the dot product of two points in that space would suffice; this is also named kernel trick and is used widely in other fields. Using an appropriate kernel, SVM is usually capable of yielding high classification performances [62-63].

- **RF**

Random forests (RFs) are an ensemble learning-based method proposed to make the decision trees robust to outliers [64]. The basic idea is to train many trees and determine the final output based on voting among their outputs. To make the final results robust, each tree is trained only on a fraction of the data, and also each tree sees a fraction of all features. The picked ratio for both of these is the square root of the available number.

- **LDA**

Used as a tool for dimension reduction, classification, and data visualization [65]. It is simple and robust and yields interpretable classification results [65]. It works by dividing the data space into $K$ disjoint regions that represent all the classes; then, in the testing phase, the label is determined by finding the region in which the data belongs. LDA can be used as the first benchmarking baseline before other, more complicated ones are employed for real-world classification problems [65].

- **KNN**

This classifier is among the simplest yet efficient algorithms; its main idea is to assign the label of each data point based on the label of those closest [66]. Consequently, there is no training phase; however, for each test subject, the distance to all training points must be calculated, which scales with the size of the dataset; thus, this method is not applicable on enormous datasets. After finding the closest points, the final label is determined using a voting scheme [66].

Table 2: Automated diagnosis of ASD with MRI neuroimaging modalities using ML methods.

| Reference | Dataset | Number of cases | Modalities | Atlas + Pipeline | Feature Extraction | Feature Selection | Classification | The best Performance criteria (%) |
|---|---|---|---|---|---|---|---|---|
| [67] | NDAR | 39 ASD | rs-fMRI | Brainnetome (BNT) Atlas | GLM Features | RFE | RF | Acc=72 |
| | | | sMRI | MNI-152 Atlas | | | | |
| [68] | ABIDE | 505 ASD, 530 HC | rs-fMRI | CC400 Atlas + CPAC Pipeline | Different Features | Nilearn | Ridge | Acc=71.98 Pre=71.53 Rec=70.89 |
| [69] | NDAR | 30 ASD, 30 HC | sMRI | NA | Cortical Path Signature Features | -- | Siamese Verification Model | Acc=87 Sen=83 Spe=90 |
| [70] | ABIDE | 103 ASD, 106 HC | rs-fMRI | AAL Atlas + DPARSF Pipeline | Graph-Theoretic Indicators (Dimensional Features) | -- | GERSVMC | Acc=96.8 |
| [71] | ABIDE | 222 ASD, 246 HC | rs-fMRI | HO Atlas + CPAC Pipeline | GARCH Model | T-Test | SVM | Acc=75.3 |
| [72] | UMCD | 51 ASD, 41 HC | DTI | NA | Graph Theory-based Features | PCA | SVM | Acc=75 Sen=81.94 Spe=70 Pre=70.42 |
| [73] | ABIDE | 250 ASD, 218 HC | rs-fMRI | AAL Atlas + CPAC Pipeline | Dimensional Feature Vectors | -- | Elastic Net | Acc=83.33 |
| [74] | Clinical | 20 ASD | sMRI | NA | GLM | Different Feature Selection Methods | RF | NA |
| | | | rs-fMRI | | | | | |
| [75] | ABIDE | 66 ASD, 66 HC | sMRI | NA | Morphological and MFN Features | RFE | SVM | Acc=78.63 Sen=80 Spe=77.27 |
| [76] | NDAR | 122 ASD, 141 HC | DTI | MNI-152 Atlas | Global and Local Feature Extraction | Signal to Noise Ratio (s2n) Filter Based Feature Ranking | SVM | Acc=71 Sen=72 Spe=70 |
| [77] | NDAR | 57 ASD, 34 HC | sMRI | NA | Morphometrical Features | -- | K-Means Clustering | NA |
| [78] | NA | 2400 ASD | Different modalities | NA | Latent Clusters | + Bayesian Information Criterion | Linear Regression (LR) | Intensity=94.29 |
| [79] | ABIDE | 175 ASD, 234 HC | rs-fMRI | AAL Atlas | Patch-based Functional Correlation Tensor (PBFCT) Features, FC Features | MSLRDA, T-Test | Multi-View Sparse Representation Classifier (MVSRC) | NA |
| [80] | NDAR | 72 ASD, 113 HC | sMRI | Desikan-Killiany (DK) Atlas | Morphological, Volumetric, and Functional Connectivity Features | -- | KNN, RF | Acc=81 Sen=84 Spe=79.2 |
| | | | rs-fMRI | | | | | |
| [81] | NA | 189 ASD, 515 HC | AQ | NA | Different Fratures | Chi-Squared Test, LASSO | LR | Acc=97.54 Sen=100 Spe=96.59 |
| [82] | UCI | 104 ASD | ASD Scan Data | NA | Different Features | Grid Search Method | RF | Acc=100 Sen=100 Spe=100 |

| Ref | Dataset | Subjects | Modality | Atlas/Pipeline | Features | Feature Selection | Classifier | Results |
|---|---|---|---|---|---|---|---|---|
| [83] | ABIDE | 392 ASD, 407 HC | rs-fMRI | DPARSF Pipeline | ICA + Different Features (Reproducible REs, NMI Values, AC Maps) | gRAICAR | K-Means Clustring | Acc=82.4 Sen=77 Spe=87 |
| [84] | ABIDE 1 | 403ASD, 468 HC | rs-fMRI | AAL Atlas + CPAC Pipeline | Dynamic Functional Connectivity (DFC) and Mean Time Series Features | MTFS-EM | SVM | Acc=76.8 Sen=72.5 Spe=79.9 |
| [85] | ABIDE | 255 ASD, 276 HC | rs-fMRI | DPARSF Pipeline | Functional Connectivity Features | RFE | SVM | Acc=90.6 Sen=90.62 Spe=90.58 |
| [86] | Clinical | 46 ASD, 39 DD (Developmental Delay) | sMRI | DK Atlas | Neuroanatomical Features (Regional Cortical Thickness, Cortical Volume, Cortical Surface Area) | -- | RF | Acc=80.9 Sen=81.3 Spe=81 AUC=88 |
| [87] | CFMRI | 46 ASD, 47 HC | Different Modalities | Johns Hopkins (JH), HO Atlas | Anatomical Variables, Cortical, Mean Diffusivity Values, Conectivity Matrices, and DTI Features | -- | Conditional Random Forest (CRF) | Acc=92.5 Sen=97.8 Spe=87.2 |
| [88] | Clinical | 24 ASD, 21 HC | sMRI | NA | Morphological Features of Subcortical Volumes | -- | LR | Acc=73.2 |
| [89] | ABIDE | 54 ASD, 57 HC | sMRI / t-fMRI | Different Atlase + DPARSF Pipeline | Regional Morphological Features | HSL-CCA, PCA | Linear SVM | Acc=81.6 F1-S =81.4 |
| [90] | NDAR | 123 ASD, 160 HC | sMRI / rs-fMRI | All Atlases | PICA (Spatial Components, Correlation Values, Power Spestral Densities) | SAE | SVM | Acc=92 Sen=93 Spe=89 |
| [91] | ABIDE 1 | 260 ASD, 308 HC | rs-fMRI | AAL Pipeline | -- | -- | Attention Based Semi-Supervised Dictionary Learning (ASSDL) Model | Acc=98.2 |
| [92] | ABIDE 1 | 250 ASD, 218 HC | rs-fMRI | AAL Atlas + CPAC Pipeline | Multi Center Domain Adaptation (MCDA) Method | -- | KNN | Acc=73.45 Sen=69.23 Spe=79.17 |
| [93] | ABIDE 1 | 155 ASD, 186 HC | sMRI | DK Atlas | Low-Order Morphological Connectivity Network (LON), Single Cell Interpretation Via Multi-Kernel Learning (SIMLR), Similarity Matrix | -- | Hypergraph Neural Network (HGNN) | Acc=75.2 |
| [94] | ABIDE | NA | sMRI / rs-fMRI | NA | GLCM | -- | ANN | NA |
| [95] | Clinincal | 30 ASD, 30 HC | t-fMRI | BNT Atlas | GLM Feature Extraction | -- | Stacked Nonnegativity Constraint Auto-Encoder (SNCAE) | Acc=75.8 Sen=74.8 Spe=76.7 |
| [96] | ABIDE 1 | 109 ASD, 144 HC | rs-fMRI | AAL, Dosenbach 160, CC 200 Atlas + DPARSF Pipeline | Sparse Low-Rank Functional Connectivity Network | Different Feature Selection Methods | SVM | Acc=81.74 Sen=71.83 Spe=89.50 |
| [97] | ABIDE 1 | 870 Subjects | rs-fMRI | AAL, multi-subject dictionary learning (MSDL) Atlas + CPAC Pipeline | ROIs Extraction, Connectivity Graphs Construction + Minimum Spanning Trees Extraction | MSTs Elimination | SVM | Acc=74,89 Sen=24,19 Spe=93,59 |
| [98] | Clinical | 30 Subjects | t-fMRI | BNT Atlas | Multi-Level GLM + GLM3 Parameters, Z-Stats Maps for All Brain Areas | RFE | RF | Acc=78 |

| Ref | Dataset | Subjects | Modality | Atlas/Pipeline | Method | Feature Selection | Classifier | Results |
|---|---|---|---|---|---|---|---|---|
| [99] | ABIDE 1 | 250 ASD, 218 HC | rs-fMRI | AAL Atlas + CPAC Pipeline | Multi-Site Adaption Framework Via Low-Rank Representation Decomposition (maLRR) Method | -- | SVM, KNN | Acc=73:44 Sen=75:79 Spe=69:52 |
| [100] | NDAR | 22 ASD, 25 HC | t-fMRI / sMRI | Proposed Atlas | GLM Analysis | -- | Stacked Autoencoder With Non-Negativity Constraint (SNCAE) | Acc=94.7 |
| [101] | ABIDE 1 / ABIDE II | 34 ASD, 34 HC / 42 ASD, 41 HC | sMRI | HO Atlas | Curvelet Transform + Coefficient Distribution Per Curvelet Sub-Band | Generalized Gaussian Distribution (GGD) | SVM | Different Results |
| [102] | ABIDE 1 | 432 ASD, 556 HC | rs-fMRI | CC200 Atlas+ DPARSF Pipeline | Graph-Theoretic Measures, Traditional FC Data | Recursive-Cluster-Elimination (RCE) | SVM | Acc= 70.1 |
| [103] | ABIDE 1 | 145 ASD, 157 HC | rs-fMRI | CC200 Atlas + CPAC Pipeline | Two-Group Cross-Localized Hidden Markov Model | Likelihood Values | SVM | Acc=74.9 |
| [104] | IMPAC | 418 ASD, 497 HC | rs-fMRI | All Atlases | Tangent-Space Embedding Metric | Permutation Feature Importance (PFI) | DenseFFwd | Acc=75.4-80.4 |
| [105] | Different Datasets | 72 ASD, 113 HC | sMRI / rs-fMRI | DK Atlas | Anatomical and Connectivity Matrix Features | -- | KNN, RF, and SVM | Acc=81 Sen=78 Spe=83.5 |
| [105] | Different Datasets | 97 ASD, 56 HC | DTI | JH Atlas | Global Features (FA, MD, AD) + Feature Mapping to Atlas + Local Feature Extraction (PDFs of Features for Each WM Area in the Atlas) | -- | KNN, RF, and SVM | Acc=81 Sen=78 Spe=83.5 |
| [106] | NAMIC | 2 ASD, 2 HC | sMRI | NA | Adaptive Independent Subspace Analysis (AISA) Method, Texture Analysis + Different Features | t-SNE | KNN | Acc=94.7 Sen=92.29 Spe=94.82 F1-S=93.56 |
| [107] | ABIDE 1 | 403 ASD, 468 HC | rs-fMRI / sMRI | NA | Eigenvalues and Topology Centralities Features | Backward Sequential Feature Selection Algorithm | LDA | Acc=77.7 |
| [108] | Clinical / ABIDE | 12 ASD, 12 HC / 12 ASD, 18 HC | rs-fMRI | NA | Group Independent Component Analysis (gICA) + Wavelet Coherence Maps Extraction | -- | SVM | Acc=86.7 Sen=91.7 Spe=83.3 |
| [109] | ABIDE 1 | 561 ASD, 521 HC | sMRI / rs-fMRI | DK, AAL Atlas + CCS Pipeline | Anatomical Feature Extraction + Functional Connectivity Analysis | -- | KNN | Different Results |
| [110] | Clinical | 36 ASD, 106 HC | sMRI | NA | Cortical Thickness, Surface Area, and Subcortical Volume Features | PCA | SVM | Different Results |
| [111] | ABIDE 1 | 155 ASD, 186 HC | sMRI | DK Atlas | Low-Order Morphological Network Construction (LON), High-Order Morphological Network Construction (HON) Features | t-SNE, K-Means Clustering | SVM | Acc=61.7 |
| [112] | Clinical | 46 ASD, 39 DD | sMRI | Talairach, DK Atlas | Regional Cortical Thickness, Cortical Volume, And Cortical Surface Area | -- | RF | Acc= 80.9 Sen= 81.3 Spe= 81 |

| Ref | Dataset | Subjects | Modality | Atlas/Pipeline | Features | Feature Selection | Classifier | Results |
|---|---|---|---|---|---|---|---|---|
| [113] | ABIDE | 54 ASD, 46 HC | rs-fMRI | AAL Atlas + DPARSF Pipeline | LON and HONs Features | LASSO | Ensemble Classifier with Multiple Linear SVMs | Acc= 81 |
| [114] | ABIDE | 160 ASD, 160 HC | rs-fMRI | HO Atlas | Functional Connectivity Matrix | CRF | SVM | Acc=65 Sen=65 Spe=65 |
| [115] | ABIDE | 61 ASD, 46 HC | rs-fMRI | AAL Atlas | Graph Theory | -- | Random SVM Cluster | Acc=96.15 |
| [116] | ABIDE | 147 ASD, 146 HC | rs-fMRI | CC200 Atlas + DPARSF Pipeline | Two Different Features Sets | -- | SVM | Acc=61.1 Sen=61.8 Spe=60 |
| [117] | ABIDE | 42 ASD, 37 HC | rs-fMRI | NA | Functional Connectivity Matrix | -- | Different Classifiers | AUC= 97.75 |
| [118] | ABIDE | 306 ASD, 350 HC | rs-fMRI | NA | Functional Connectivity Matrix | CRF | RF | Acc= 73.75 |
| [119] | ABIDE 1 | 539 ASD, 573 HC | rs-fMRI | CPAC Pipeline | Feature Extaction (All Voxels Within Grey Matter Template Mask in MNI152 Space) | -- | SVM | Acc=62 |
| [120] | UMCD | 79 Functional and 94 Structural Connectomes | rs-fMRI | NA | Graph Theory + Global, Nodal Measurements and Gender Information | Relieff Algorithm | Ensemble Learning | Acc=67 pre=0.67 Recall=70 |
| [120] | UMCD | 79 Functional and 94 Structural Connectomes | DTI | NA | Graph Theory + Global, Nodal Measurements and Gender Information | Relieff Algorithm | Ensemble Learning | Acc=68 Pre=0.73 Rec=70 |
| [121] | NDAR | 124 ASD, 139 HC | DTI | JH Atlas | Global and Local Features | Signal to Noise Ratio (S2n) Filter | SVM | Acc= 73 Sen= 70 Spe= 76 |
| [122] | ABIDE II | 31 ASD, 23 HC | rs-fMRI / DTI / sMRI | AAL Atlas | Connectivity Matrix | -- | SVM | Acc= 72.34 |
| [123] | ABIDE / Clinical | 126 ASD, 126 HC / 42 ASD, 30 HC | rs-fMRI | NA | Functional Connectivity Matrix | CRF | SVM | Acc > 90 |
| [124] | ABIDE | 167 ASD, 205 HC | rs-fMRI | CCS Pipeline | Functional Connectivity Matrix | -- | SVM | Different Results |
| [125] | ABIDE 1 | 403 ASD, 465 HC | rs-fMRI | HO Atlas + CPAC Pipeline | sFC, dFC, and Haralick Texture Features | -- | SVM | -- |
| [126] | ABIDE | Whole Dataset | rs-fMRI | AAL Atlas + DPARSF Pipeline | Pearson Correlation Coefficient, Graph Measures, and Mean Intensities Features | -- | Adaboost | Acc=66.08 |
| [127] | Clinical | 46 ASD, 47 HC | sMRI / DWI / rs-fMRI | JH Atlas / HO Atlas | Functional Connectivity Matrix Features | -- | CRF | Acc=92.5 Sen=97.8 Spe=87.2 |
| [128] | Clinical / ABIDE | 19 ASD / 64 ASD | t-fMRI / rs-fMRI | NA | Elastic Net Regression | -- | RF | NA |
| [129] | ABIDE 1 | 816 Subjects | rs-fMRI | AAL Atlas + CPAC Pipeline | Graph Theoretical Metrics | Sequential Forward Floating Algorithm | SVM | Acc=95 Sen=97 Spe=91 |
| [130] | ABIDE 1 / ABIDE II | 119 ASD, 116 HC / 97 ASD, 117 HC | rs-fMRI | AAL, CC200 Atlas + DPARSF Pipeline | Community Pattern Quality Metrics Features | -- | LDA, KNN | Acc= 75 Prec= 76.07 Rec= 71.67 |
| [131] | Clinical | 64 ASD, 66 ADHD, 28 HC | rs-fMRI | NA | 43 Executive Functions (EF) | -- | Functional Random Forest (FRF) | Different Results |

| Ref | Dataset | Subjects | Modality | Atlas/Pipeline | Features | Feature Selection | Classifier | Results |
|---|---|---|---|---|---|---|---|---|
| [132] | Clinical | 29 ASD, 31 HC | sMRI | Different Atlas | Graph Theory + Different Features | Statistical Analysis | SVM | Acc= 92 |
|  |  | 20 ASD, 20 HC | t-fMRI |  |  |  |  |  |
| [133] | ABIDE 1 | 21 ASD, 26 HC | rs-fMRI | AAL Atlas + DPARSF Pipeline | Fast Entropy Algorithm + Important Entropy | -- | SVM | AUC= 62 |
| [134] | ABIDE 1 | 59 ASD, 46 HC | rs-fMRI | AAL Atlas + DPARSF Pipeline | Function Connectivity + Minimum Spanning Tree (MST) | -- | SVM | Acc=86.7 Sen=87.5 Spec=85.7 |
| [135] | ABIDE 1 | 437 ASD, 511 HC | sMRI | -- | Computing the Brain Asymmetry with The BrainPrint + Asymmetry Values | -- | LR Models | NA |
| [136] | Clinical | 14 ASD, 33 HC | MRI, DTI | DK Atlas | Different Features | -- | Naïve Bayes, RF, SVM, NN | Acc= 75.3 Sen= 51.4 Spec= 97.0 |
| [137] | ABIDE | 45 ASD, 47 HC | rs-fMRI | AAL Atlas | Modified Weighted Clustering Coefficients | t-Test and SVM-RFE | Multi-Kernel Fusion SVM | Acc=79.35 Sen=82.22 Spec=76.60 |
| [138] | ABIDE I | 505 ASD, 530 HC | rs-fMRI | CC200 Atlas + CPAC Pipeline | Functional Connectivity | Graph-Based Feature Selection | MMoE Model | Acc=68.7 Sen= 68.9 Spec= 68.6 |
| [139] | ABIDE, | 86 ASD, 83 ADHD, 125 HC | sMRI, rs-fMRI | DK Atlas | Functional Connectivity | Univariate t-Test and Multivariate SVM-RFE | SVM | Acc=76.3 Sen= 79.2 Spec= 63.9 |
| [140] | ABIDE | 24 ASD, 35 HC | rs-fMRI | AAL Atlas | Mutual Connectivity Analysis with Local Models (MCA-LM) | Kendall's τ Coefficient | RF and AdaBoost | Acc= 81 |
| [141] | ABIDE II | 23 ASD, 15 HC | rs-fMRI | AAL Atlas + AFNI Pipeline | Functional Connectivity | ANOVA F-Score | SVM | Acc=80.76 |
| [142] | ABIDE 1 | 74 ASD, 74 HC | fMRI | DPARSF, CCS Pipeline | Bag-of-Feature (BoF) Extraction | -- | SVM | Acc=81 Sen=81 Spec=86 |
| [143] | ABIDE | 70 ASD, 74 HC | fMRI | NA | Functional Connectivity | Elastic SCAD SVM | SVM | Acc=90.85 Sen=90.86 Spec=90.90 |
| [144] | ABIDE | 250 ASD, 218 HC | rs-fMRI | AAL Atlas + CPAC Pipeline | Functional Connectivity + Low-Rank Representation Decomposition (maLRR) | -- | KNN, SVM | Acc= 73.44 Sen= 75.79 Spec= 69.52 |
| [145] | ABIDE | 399 ASD, 472 HC | rs-fMRI | CC200 Atlas + CPAC Pipeline | Feature Extraction (Static FC, Demographic Information, Haralick Texture Features, Kullback-Leibler Divergence) | Feature Selection Algorithms (RFE-CBR, LLCFS, InfFS, mRMR, Laplacian Score) | SVM, KNN, LDA, Ensemble Trees | Acc=72.5 Sen=94 Spec=64.7 |
| [146] | ABIDE | 408 ASD, 476 HC | rs-fMRI | CPAC Atlas | 5 Methods for Functional Connectivity Matrix Construction | 6 Feature Extraction/Selection Approaches | 9 Classifiers | -- |
| [147] | Clinical | 30 Pairs of Biological Siblings | rs-fMRI | Social Brain Connectome Atlas | Functional Connectivity | Sparse LR (SLR) | Bootstrapping Approach | Acc=75 Sen= 76.67 Spec= 73.33 |
| [148] | Clinical | 26 ASD, 24 CAS, 18 HC | sMRI | -- | Feature Extraction | Statistical Analysis | SVM | AUC= 73 |
| [149] | Clinical | 15 ASD, 15 HC | Task-fMRI | -- | Functional Connectivity + Effective Connectivity | -- | RCE-SVM | Acc= 95.9 Sen= 96.9 |

| Ref | Dataset | Subjects | Modality | Atlas/Pipeline | Features | Feature Selection | Classifier | Results |
|---|---|---|---|---|---|---|---|---|
| | | | | | | | | Spec= 94.8 |
| [150] | ABIDE 1 | -- | rs-fMRI | CC200, AAL Atlas + CPAC Pipeline | Graph Extraction + Feature Extraction | PCA | MLP | Different Results |
| [151] | ABIDE | 119 ASD, 116 HC | rs-fMRI | AAL Atlas + DPARSF Pipeline | Resting-State Functional Network Community Pattern Analysis | RFE | LDA | Acc= 74.86 Prec= 76.07 Recall= 71.67 |
| [152] | ABIDE | 42 ASD, 37 HC | rs-fMRI | -- | Functional Connectivity + Joint Symmetrical Non-Negative Matrix Factorization (JSNMF) | -- | SVM | AUC=97.75 |
| [153] | ABIDE | 245 ASD, 272 NC | rs-fMRI | DPARSF Pipeline | Different Features | NAG-FS | SVM | Acc=65.03 |
| [154] | ABIDE 1 | 201 ASD, 251 HC | rs-fMRI | AAL Atlas + CPAC Pipeline | Graph Construction + Graph Signal Processing (GSP) | Fukunaga-Koontz Transform (FKT) | DT | Acc=75 |
| [155] | ABIDE 1 ABIDE II | 133 ASD, 203 HC 60 ASD, 89 HC | rs-fMRI, sMRI | -- | Functional Connectivity | Statistical Analysis | Sparse LR | Acc=82.14 Sen=79.70 Spec=83.74 |
| [156] | ABIDE II | 24 ASD, 35 HC | rs-fMRI | AAL Atlas | large-scale Extended Granger Causality (lsXGC) | Kendall's Tau rank correlation coefficient | SVM | Acc= 79 |
| [297] | Clinical | 15 ASD, 15 HC | fMRI | NA | Functional Connectivity, Effective Connectivity and Fractional anisotropy (FA) From DTI, Behavioral Scores | Recursive Cluster Elimination | SVM | Acc=95.9 |
| [298] | Clinical | 22 ASD, 16 HC | MRI | Cortical Atlas | Thickness and Volume-Based Features | Surface-Based Morphometry | Different Cassifiers(SVM,FT, LMT) | Acc=87 Sen=95 Spe=75 |
| [299] | Clinical | 22 ASD, 22 HC | MRI | NA | GLM, Different Features | RFE-SVM | SVM | Spe=86 Sen=88 |
| [300] | ABIDE | 126 ASD, 126 HC | rs-fMRI | NA | Pearson Correlation Matrix, Connectivity Measures | PSO-SVM | SVM -RFE | Acc=66 Sen=60 Spe=72 |
| [301] | ABIDE | 24 ASD, 24 HC | sMRI | NA | Multivariate Statistical Pattern, Morphological Feature | NA | SVM | Acc = 80 |
| [302] | Clinical | 45 ASD, 30 HC | DTI | EVE | FA (Fractional Anisotropy), MD Mean diffusivity, Anatomical ROI's | Signal-To-Noise (s2n) Ratio Coefficient Filter | SVM | Spe=84 Sen=74 |
| [303] | Clinical | 81 ASD, 50 HC | MRI | NA | Feature Extraction (Voxelwise Tissue Density Maps For GM, WM And ventricles (VN)) | Welch's T-Test | SVM | Acc=73.28 Sen=71.6 Spe=76 |
| [304] | Clinical | 13 ASD,15 HC | fMRI | NA | Functional ROIs, Functional Connectivity, Seed-Based Connectivity | T-Test | Logistic regression | Acc > 96.3 |
| [305] | Clinical | 23ASD,22 HC | MRI | NA | Orientation Invariant Features of Each ROI's Mean FOD | PCA | SVM | Acc=77 |
| [306] | Clinical | 76 ASD,76 HC | sMRI | NA | Sequences Of The Intensity Values Of The GM Segments | SVM-RFE | SVM | Sen=82 Spe=80 |
| [307] | Clinical | 15 ASD, 15 HC | Task-fMRI | NA | Functional Connectivity, Effective Connectivity | NA | RCE-SVM | Acc= 95.9 Sen= 96.9 Spec= 94.8 |

| Ref | Dataset | Subjects | Modality | Atlas | Features | Feature Selection | Classifier | Results |
|---|---|---|---|---|---|---|---|---|
| [308] | Clinical | 20 ASD, 20 HC | MRI | NA | Morphological Parameters Including Volumetric and Geometric Features | NA | SVM | Sen=90 Spe=80 |
| [309] | Clinical | 10 ASD, 10 HC | DTI | JHU-DTI-MNI | Brain Connectivity Network | Network Regularized SVM-RFE | SVM | Acc=100 |
| [310] | Clinical | 31 Klinefelter syndrome, 8 XYY Syndrome 75 HC | sMRI | NA | Statistical Parametric Mapping (Grey Matter Volume (TGMV) A Volume (TWMV) Measures) | RFE | SVM | NA |
| [311] | Clinical, ABIDE | 79 ASD, 105 HC | MRI | NA | Voxel Locations of VBM Detected Brain Region | T-test | PBL-McRBFN | Acc (Mean)=70 Sen(Mean)=53 Spe(Mean)=72 |
| [312] | Clinical | 82 ASD, 84 HC | sMRI | NA | Inter-Regional Thickness Correlation (IRTC) Using Pearson Correlation Between the Cortical Thicknesses of Each Region. | NA | Support Vector Reression | NA |
| [313] | Clinical | DTI Data: 5 b0 iImages, followed by 30 Diffusion Weigted Images, Child Control dataset | fMRI / DTI | Brodmann | Fiber Connectivity Feature, ROIs Extraction, Functional Connectivity Information | NA | mv-EM | Max Percent Error: mv-EM: 8.55 |
| [314] | Clinical | 21 ASD, 21HC | fMRI | NA | Neural Substrates And Inter-Individual Functional Connectivity | T-test | NA | Acc=74.2∓1.9 |
| [315] | BLSA | 17 MCI (mild cognitive impairment) | MRI | NA | Tissue Density Maps, Top-Ranked Features Wavelet Decomposition Level | Wavelet-Based Data Compression | JointMMCC | Different Results |
| [316] | Clinical | 38 ASD, 38 HC | sMRI | NA | Volumetric Variables (GM, WM, CSF, TIV), | SVM-RFE, T-test | SVM | AUC= 80 |
| [317] | Clinical | 13 ASD | MRI | NA | Regional Cortical Thicknesses And Volumes | NA | Three Decision-Tree-Based Models, SVM, logistic Model Tree | Acc > 80 Spe >34 Sen > 92 |
| [318] | ABIDE | 447 ASD, 517 HC | rs-fMRI | NA | Functional Connectivity From a lattice of ROIs Covering The Gray Matter | NA | leave-one-out | Acc=60 Spe=58 Sen=62 |
| [319] | Clinical | 22 ASD, 16 HC | MRI | NA | Using Surface-based morphometry For Cortical Features (Average thickness, Mean Curvature, Gaussian curvature, Folding index, Curvature index) | NA | SVM,FT,LMT | Acc (SVM)=74 Acc(FT)=76 Acc(LMT)=76 |
| [320] | Clinical | 76 ASD, 76 HC | sMRI | NA | GM Volumes | RFE | SVM | AUC=82 |
| [321] | Clinical | 41 ASD, 40 HC | sMRI | NA | Regional Features | -- | SVM | AUC=81 |
| [3222] | ABIDE | 505 ASD, 530 Neurotypical Subjects | rs-fMRI | NA | Spatial Feature based Detection Method (SFM) (Mean Connectivity Matrices, Discriminative Log-variance Features) | Feature Selection Based on top m Signals | SVM | Acc=95 |
| [323] | Clinical | 41 ASD, 40 HC | sMRI | NA | ROI Features | -- | SVM | AUC=74 |

| Ref | Dataset | Subjects | Modality | Preprocessing | Features | Feature Selection | Classifier | Results |
|---|---|---|---|---|---|---|---|---|
| [324] | Clinical | 35 ASD, 51 TD, 39 No Known Neuropsychiatric Disorders | fMRI | NA | Individual Difference Measures in BOLD Signals | -- | LR | Sen=63.64 Spe=73.68 |
| [325] | ABIDE | 112 ASD, 128 HC | rs-fMRI | NA | Functional Connectivity Values | F-score Method | SVM | Acc=79.17 |
| [326] | NDAR | 58 ASD, 59 HC | sMRI | NA | Regional and Interregional Morphological Features | T-Test / mRMR | SVM | Acc=96.27 AUC=99.52 |
| [327] | ABIDE | 127 ASD, 153 TD | sMRI / rs-fMRI | NA | Quantitative Imaging Features (Regional Gray Matter and Cortical Thickness Volumes0 | mRMR | SVM | Acc=70 |

## 5. Challenges in detecting ASD with MRI neuroimaging modalities and AI techniques

This section introduces the challenges facing in ASD detection from MRI neuroimaging modalities and AI techniques. The challenges mentioned in this section cover dataset limitations, lack of access to multimodal datasets, AI techniques, and suitable hardware resources. They are briefly described below.

- **Unavailable MRI neuroimaging datasets with different ASD patient**

All datasets available involve two classes of ASD and control fMRI or sMRI modalities. However, there are different types of ASD, and this poses a serious obstacle for researchers in AI wishing to develop systems that can detect different types of disorders. Datasets with different types of ASD can help pave the way for accurate diagnosis of various types of ASD.

- **Unavailable multi-modalities datasets for ASD diagnosis**

In medical research, specialists have shown that using neuroimaging multimodalities can effectively improve brain disorders diagnosis. Neuroimaging modality fusion is one of the newest methods for diagnosing brain disorders such as ASD [355], SZ [356], and ADHD [357]. Physicians usually use MRI data with other neuroimaging modalities to diagnose brain disorders. To diagnose neurological and mental disorders, fMRI-MEG [358], MRI-PET [359], and EEG-fMRI [360] are the most important multimodalities. Unfortunately, the neuroimaging multimodalities datasets are not available for studies on ASD diagnosis. Such datasets might lead to practical and interesting studies in ASD diagnosis.

- **Challenges in AI algorithms in diagnosing ASD**

CADS based on ML algorithms are highly time-consuming and complex to design. However, if the appropriate algorithms are selected, the resulting CADS can accurately diagnose ASD. DL methods automatically perform the steps from feature extraction to classification. By using intelligent feature extraction, DL eliminates the need for supervision on features, which may reduce the performance of a CADS based on DL compared to ML. Therefore, when ML methods are combined with DL, promising results can be obtained in CADS for diagnosis of ASD.

- **Challenges in hardware's**

The lack of access to appropriate hardware resources is another problem encountered by researchers in the field of automated ASD detection. ASD detection datasets that are available publicly, such as ABIDE, have a lot of data; this poses many challenges for the storage and processing of these datasets on ordinary computers. In contrast, research in CAD implementation on cloud servers has not been seriously conducted to eliminate hardware resource problems. As a result, cloud servers are not yet extensively used for data storage and processing. Recently, some DL models called deep compact CNN models have been introduced to be implemented on hardware systems with limited resources. Deep compact-size CNN models require fewer hardware sources than other CNN methods [361-362]. Some deep compact-size CNN methods include FBNetV3 [363], MobileNet [364], and TinyNet [365].

## 6. Discussion

This paper presents and compares the research about automated ASD detection with MRI neuroimaging modalities and AI methods. First, this section comprehensively compares the conducted studies on ASD detection using ML and DL techniques. In subsection one, the number of studies conducted annually in the ASD detection from MRI neuroimaging modalities using different ML and DL techniques are presented. In subsection two, the MRI datasets employed in studies on the automated diagnosis of ASD using ML and DL techniques are compared. In subsection three, the number of MRI studies conducted annually on ASD detection from MRI neuroimaging modalities are discussed. The employed atlases in ML and DL studies for ASD detection are introduced in subsection four. Section five discusses MRI pipeline techniques in the diagnosis of ASD research using ML and DL methods. Ultimately, different classification algorithms for ML and DL-based diagnosis of ASD are compared.

- **Comparison between the numbers of papers published each year for ML and DL research**

This section presents the number of published papers annually on ASD detection using AI techniques. Studies on the ASD detection from MRI modalities and ML and DL techniques began in 2017. Table (2) represents the papers on ASD detection in MRI neuroimaging modalities using ML methods. In addition, articles in ASD detection in MRI neuroimaging modalities using DL techniques are introduced in Appendix A. Figure (4) illustrates the number of papers published annually on ML and DL techniques for ASD detection.

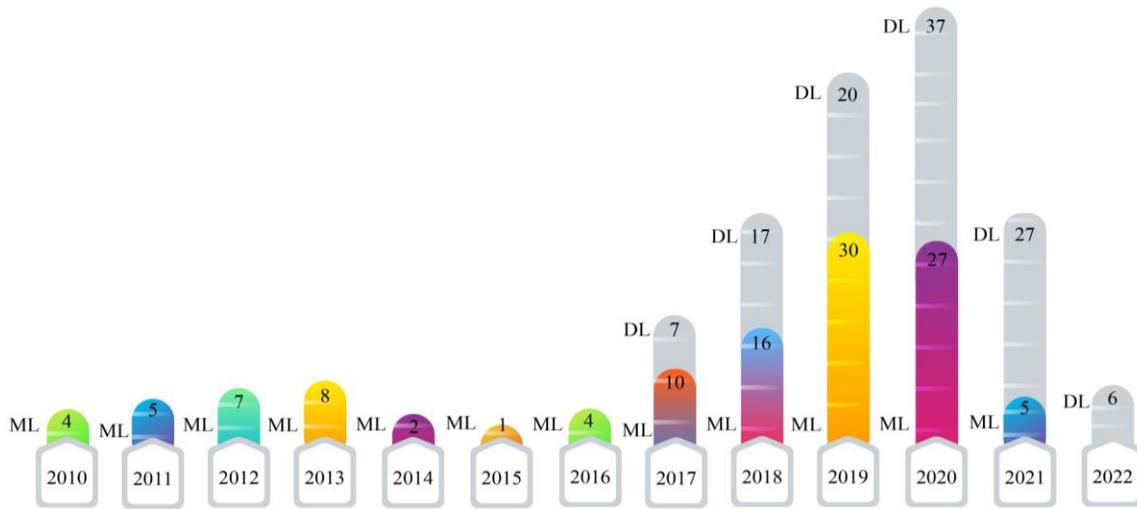

Fig. 4. Shows the number of papers published in ASD detection using ML and DL methods.

As demonstrated in recent years, researchers' interest in using DL architectures has significantly grown compared to ML techniques. According to Fig. (4), DL models are used more in studies on the automated diagnosis of ASD with MRI modalities than ML models. Therefore, implementation of CADS based on DL techniques is promising for developing applied software for ASD detection with MRI neuroimaging modalities in the future. For automated diagnosis of ASD with MRI modalities, various datasets are proposed in ABIDE. Besides, various toolboxes are available for the implementation of different DL models. These reasons are the foundation for many studies on the automated diagnosis of ASD using DL models.

- **Comparison between the numbers of datasets used in the ML and DL research**

As stated in the neuroimaging modalities section, limited datasets are accessible. ABIDE is the most important dataset available in this field, which includes two datasets, ABIDE I and ABIDE II. Figure (5) demonstrates the types of datasets employed in the automated ASD diagnostic research using DL and ML techniques.

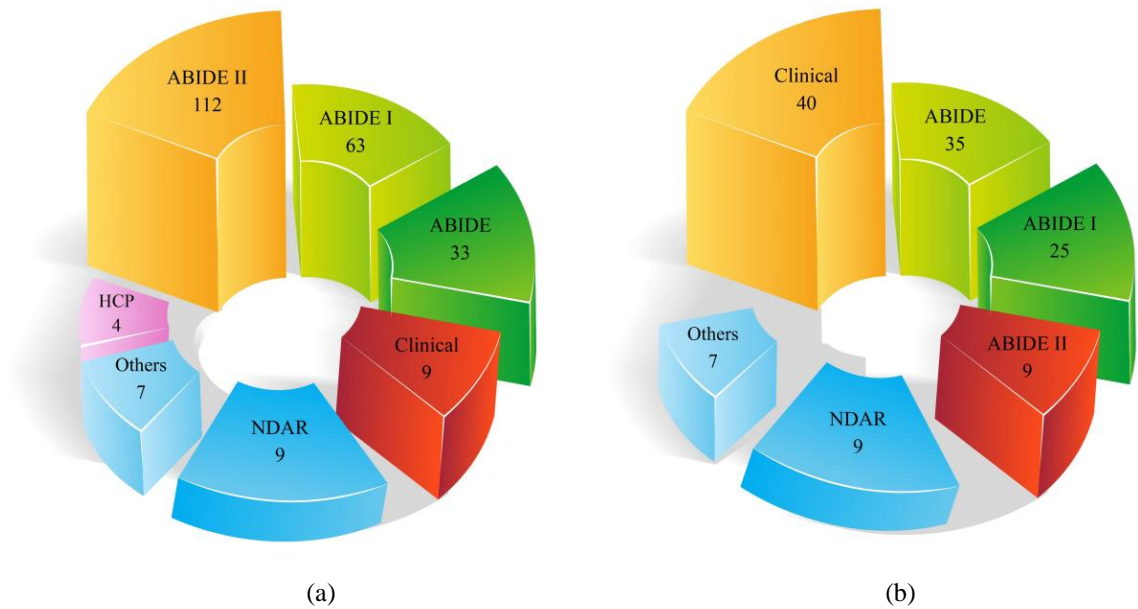

Fig. 5. Number of datasets used for automated ASD detection. a) DL and b) ML methods.

It can be noted from Fig. (5a) and (5b) that a greater number of ABIDE datasets are employed in studies on the automated diagnosis of ASD. The major reason for the wide use of this dataset in various studies on the automated diagnosis of ASD is the availability of many subjects and different MRI modalities.

- **Comparison between the numbers of neuroimaging modalities used in the ML and DL research**

The different structural and functional MRI neuroimaging modalities and ML and DL methods, play an essential role in automated ASD detection. In Table (2) reports studies on the automated ASD detection using ML techniques and different MRI neuroimaging modalities have been presented. Moreover, Table (3) discusses the ASD detection using DL techniques. Figures (6a), and (6b) describes the annual research carried out to detect automated ASD using sMRI and fMRI neuroimaging modalities.

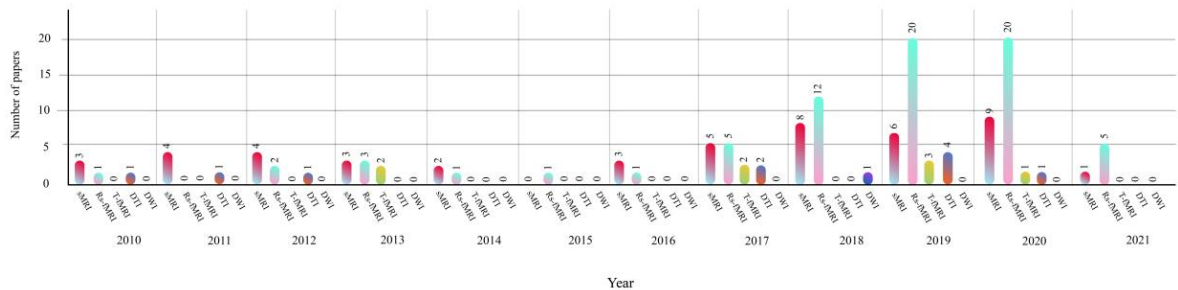

Fig. 6a. Shows the number of MRI neuroimaging modalities used in the CADS based on ML methods.

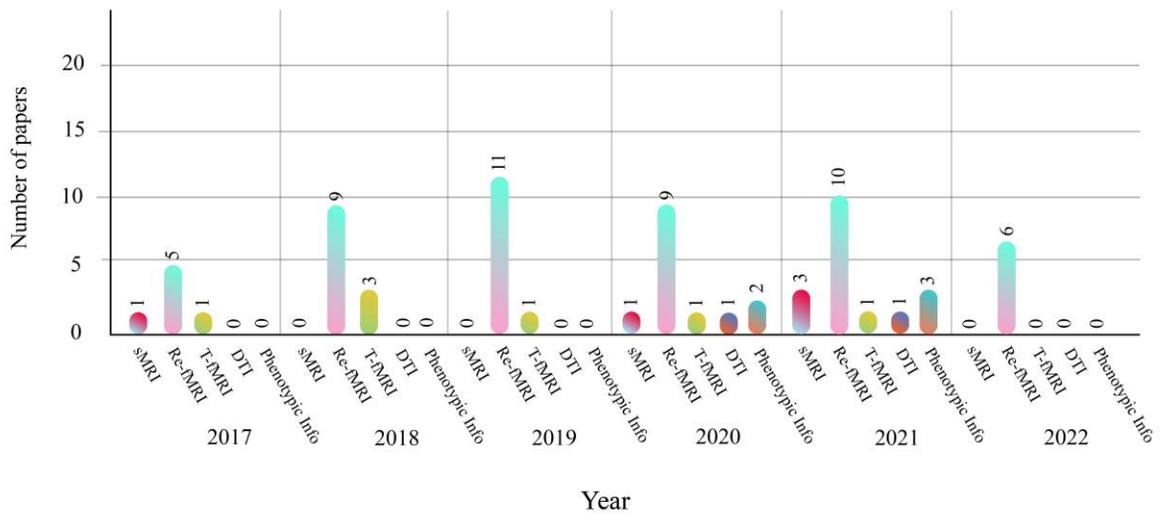

Fig. 6b. Shows the number of MRI neuroimaging modalities used in the CADS based on DL methods.

As shown in Figures (6a) and (6b), the rs-fMRI modalities are most used in studies on ASD detection using ML and DL methods. As mentioned earlier, ASD is a neurological disorder that negatively affects brain function. Accordingly, researchers have used rs-fMRI modalities most widely in studies on ASD detection using AI methods.

- **Comparison between the numbers of Atlases used in the ML and DL research**

In another part of Tables (2) and (3), the types of Atlases for MRI neuroimaging modalities have been provided. Atlases are considered an important preprocessing step discussed in part of this section. The number of atlases employed in ML and DL research are described in Figure (7).

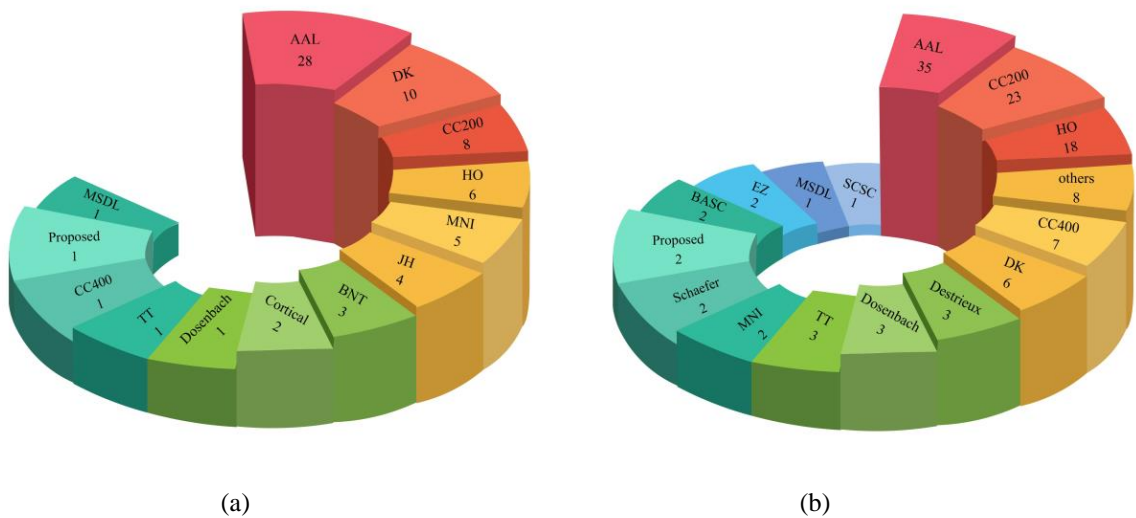

(a)          (b)

Fig. 7. Number of Atlas used for ASD Detection. a) ML and b) DL methods

As shown in Fig. (7a) and (7b), the AAL atlas is most used in studies for ASD detection in MRI neuroimaging modalities using AI methods.

- **Comparison between the numbers of pipelines used in the ML and DL researches**

Pipelines play a significant role in preprocessing of MRI modalities. The pipelines employed in ASD data preprocessing are presented in Tables (2) and (3). The number of pipelines utilized in DL and ML research is shown in Figure (8). The results of the studies reveal that the CPAC pipeline is the most widely used.

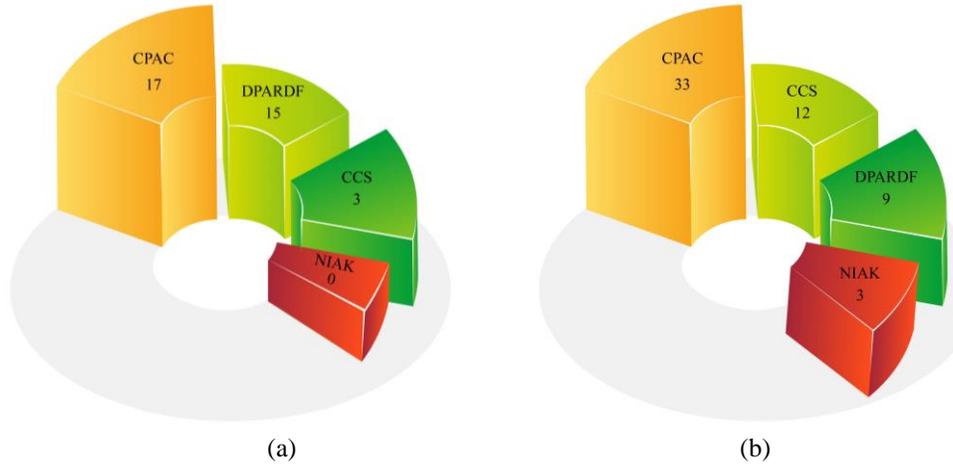

(a)          (b)

Fig. 8. Number of pipelines used for ASD Detection: a) ML and b) DL methods.

- **Comparison between the numbers of classification methods in the ML and DL research**

Classification is the last step of CADS with ML or DL methods. So far, various classification methods have been proposed in ML and DL, which are presented in Tables (2) and (3). The types of classification algorithms applied in CADS using DL and ML are depicted in Figure (9). As shown in these figures, (9a) and (9b), it may be noted that the Softmax method is most used in DL architectures. In addition, compared to other classification methods, SVM is the most widely applied in ML methods.

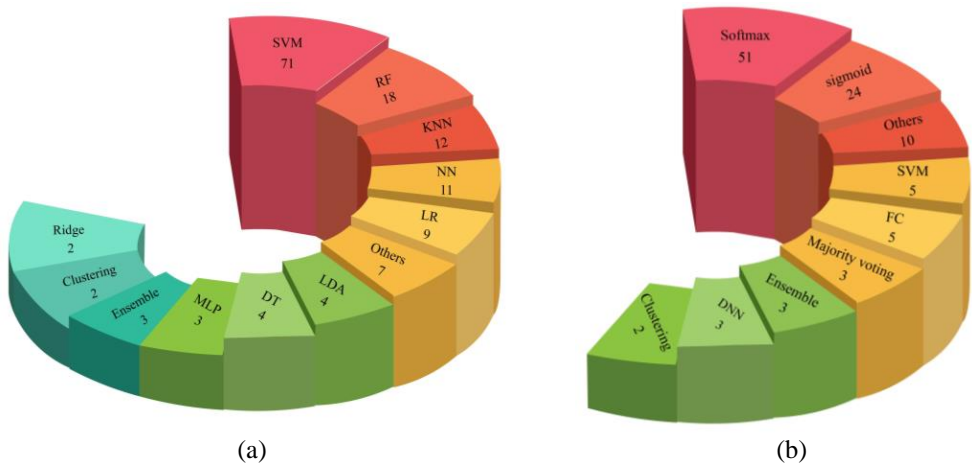

(a)          (b)

Fig. 9. Number of classifiers used in CADS for ASD detection: a) ML and b) DL methods.

## 7. Future Works

Lack of access to huge public datasets with various ASD disorders researchers is a big challenge. As mentioned in the introduction, autism has different types [2], and the availability of datasets containing different types of ASD is of paramount importance for researchers. Hence, presenting MRI datasets of different types of autism disorder need to be addressed in future works. These datasets help researchers conduct more studies and compare their studies with other researchers on the automated diagnosis of ASD. As mentioned in previous sections, ABIDE is a free dataset available for researchers and consists of different cases and MRI modalities of ASD patients. But it does not have a large number of cases of DTI modalities for the diagnosis of ASD. DTI modality is one of the popular methods in ASD detection. Publicly providing more datasets of this type of modality could increase research in the ASD diagnosis field using DTI modality.

Another future work is to provide multimodal datasets, such as fMRI-EEG, for the diagnosis of ASD. In clinical studies [158], it has been indicated that using multimodal neuroimaging, such as fMRI-EEG, plays a pivotal role in diagnosing ASD. Providing datasets with combined modalities paves the way for new studies on the diagnosis of ASD using different AI methods.

Automated diagnosis of ASD with MRI using ML techniques can be the other future work. Various methods have been proposed for feature extraction from MRI data for the diagnosis of ASD, which are summarized in Table (2). According to Table (2), fuzzy-based feature extraction techniques have not been used in the diagnosis of ASD, and they can be introduced in future work. Fuzzy techniques are important in medical applications and allow researchers to develop software close to human logic [159-164]. Hence, providing graph models based on fuzzy theory can be addressed in the future, leading to the accurate diagnosis of ASD with MRI modalities. Connectivity techniques are an essential feature extraction method for structural and functional neuroimaging modalities [165-170]. Proposing new feature extraction methods based on connectivity for structural and functional neuroimaging modalities is also another field for future works. Table (2) also indicates classification algorithms. In this section, fuzzy type 1 and 2 techniques can be used for data classification as future work on the diagnosis of ASD [171-173]. Furthermore, in the future, graph theory-based classification methods can also be used to increase the performance of the CADS for automated diagnosis of ASD [174-175].

In Appendix (A), different studies on the automated diagnosis of ASD using MRI modalities and DL techniques is presented. It may be noted that conducted studies have used standard DL methods to diagnose ASD. In future works, graph theory [176-177], representation learning [178-179], zero-shot learning [180], Q-learning [181], attention learning [182], and advanced models of adversarial networks [183-184] can be used for the automated diagnosis of ASD with MRI modalities.

Feature fusion technique is a new field in diagnosing different diseases, and many studies are being conducted in this field [185-190]. The DL features can be extracted from MRI images for automated ASD detection. Ultimately, ML and DL features can be used to obtain high performance in the automated diagnosis of ASD.

## 8. Conclusion

ASD is a neurological disorder with unknown symptoms that begins in childhood and cause problems in communication, social relationships, perception processing, and repetitive behaviors. In few studies, physicians have stated that ASD usually occurs due to genetic mutations or the inability of the fetus's brain cells to obey regular growth patterns during the first steps [1-5].

Physicians use different ASD detection methods, among which different neuroimaging modalities are of paramount importance. Among different types of neuroimaging modalities, MRI-based functional and structural modalities are mostly used to diagnose ASD. sMRI and fMRI provide physicians with important information on the structure and function of the brain, respectively. Accurate diagnosis of ASD from sMRI and fMRI is sometimes time-consuming and challenging. Moreover, factors such as tiredness or different noises in MRI modalities may lead to clinicians' wrong diagnosis of ASD.

For this purpose, many studies are being conducted on the automated diagnosis of ASD using AI techniques, aiming to increase the performance of automated diagnosis of ASD. In general, studies on the automated diagnosis of ASD from MRI modalities using AI cover ML and DL methods. In few papers, researchers have conducted a review study in ASD detection based on DL [6] and ML [191-196] methods with different neuroimaging modalities.

This work is a comprehensive review of studies conducted on ASD detection in different MRI neuroimaging modalities using AI methods. First, AI-based CADS for ASD detection from different MRI neuroimaging modalities was introduced. Then, the steps of the CADS based on ML algorithms for automated ASD detection in MRI neuroimaging modalities were studied. Also, in this section, papers on the automated ASD detection in MRI neuroimaging modalities using ML methods are summarized in Table (2). Previously, some authors of this study previously published a review paper about automatic ASD detection in different neuroimaging modalities using DL techniques [6], which is summarized in Table (3). In another section, the most critical challenges in ASD detection in MRI neuroimaging modalities and AI methods were presented. Also, this section studied the most important challenges in the automated diagnosis of ASD using MRI modalities and AI techniques. The most important challenges in the diagnosis of ASD are the lack of access to public datasets with different MRI modalities, multimodal datasets, such as fMRI-EEG, AI algorithms, and hardware resources.

In the discussion section, first, the number of published annual papers he ASD detection using ML methods and DL techniques were discussed. Then, the number of datasets used in ML and DL studies was presented. In addition, the number of different MRI neuroimaging modalities with ML and DL methods used in annual studies in ML and DL was also indicated. Also, a comparison was made between different atlases used in MRI neuroimaging preprocessing for ASD detection. In another subsection, the number of pipelines in the preprocessing step of the MRI neuroimaging modalities for CADS based on various AI methods is also examined and compared. Finally, the number of classifier algorithms used in ML and DL studies for ASD detection was discussed.

In section 5, the future works for ASD detection in MRI neuroimaging modalities and AI methods were addressed. In this section, future works on MRI datasets for the diagnosis of ASD were first discussed. Then, future works on the diagnosis of ASD using AI techniques were addressed. Besides, future works on the automated diagnosis of ASD with MRI modalities were introduced. The final section also recommended the idea of using feature fusion for the diagnosis of ASD with MRI modalities in future works. Studies on ASD detection using AI techniques indicate that researchers will use the proposed methods in the future. The proposed methods are promising in developing real software for ASD detection using MRI modalities and help clinicians quickly diagnose ASD in the early stage.

**Appendix A:**

Table 3: Automated diagnosis of ASD with MRI neuroimaging modalities using DL methods.

| Work | Datasets | Neuroimaging Modalities | Image Atlas + Pipeline | Details for Deep Learning Models ||||| 
|---|---|---|---|---|---|---|---|---|
| | | | | Architecture | Layers | Optimizer | Loss Function | Classifier |
| [197] | Clinical | T-fMRI<br>Residual fMRI | MNI152 Atlas | 2CC3D | CNN (6) + Pooling (4) + FC (2) | -- | BCE | MV |
| [198] | Clinical | T-fMRI | AAL Atlas | 2CC3D | CNN (6) + Pooling (4) + FC (3) | -- | -- | Sigmoid |
| [199] | HCP | T-fMRI<br>rs-fMRI | -- | 3D-CNN | CNN (2) + LReLU + Pooling (1) + FC (1) | SGD | MNLL | Softmax |
| [200] | Clinical | T-fMRI | AAL Atlas | LSTM | LSTM (1) + Pooling (1) + FC (3) | Adadelta | MSE | Sigmoid |
| [201] | Different Datasets | T-fMRI<br>rs-fMRI<br>Phenotypic Info | AAL Atlas | 2D-CNN | CNN (2) + ReLU + BN (4) + FC (3) | Adam | -- | Softmax |
| [202] | Clinical | T-fMRI | AAL Atlas | 2CC3D | CNN (6) + Pooling (4) + FC (2) | -- | -- | Sigmoid |
| [202] | ABIDE-I | rs-fMRI | AAL Atlas | 2CC3D | CNN (6) + Pooling (4) + FC (2) | -- | -- | Sigmoid |
| [203] | ABIDE 1<br>ABIDE II | rs-fMRI | All Atlases + CPAC Pipeline | 3D-CNN | CNN (2) + ELU + Pooling (2) + FC (2) | SGD | -- | Sigmoid |
| [204] | ABIDE 1 | rs-fMRI | CC-200 and AAL Atlas + CPAC Pipeline | AE | Standard AE with Tanh Activation | -- | MSE<br>BCE | SLP |
| [205] | ABIDE 1 | rs-fMRI | HO Atlas + CPAC Pipeline | G-CNNs | Proposed G-CNN with 3 Layer CNN | Adam | -- | Softmax |
| [206] | ABIDE 1 | rs-fMRI | AAL Atlas + CCS Pipeline | BrainNet | Element-Wise (1) + E2E (2) + E2N (1) + N2G (1) + FC (3) + Leaky ReLU + Tanh | Adam | Proposed Loss function | Softmax |
| [207] | ABIDE 1 | rs-fMRI | AAL Atlas | DAE | Standard DAE | -- | Proposed Loss function | -- |
| [208] | ABIDE | rs-fMRI | NA | LeNet-5 | Standard LeNet-5 Architecture | -- | -- | Softmax |
| [209] | ABIDE 1 | rs-fMRI | AAL Atlas + NIAK Pipeline | SAEs | SAE with LSF Activation | LBFGS | -- | Softmax |
| [210] | ABIDE 1<br>ABIDE-II<br>ABIDE 1 + II | rs-fMRI | NA | MCNNE | CNN (3) + ReLU + Pooling (3) + FC (1) | Adam<br>Adamax | BCE | Binary SR |
| [211] | ABIDE | rs-fMRI | All Atlases + CPAC Pipeline | 3D-CNN | CNN (2) + ELU + Pooling (2) + FC (3) | SGD<br>Adam | BCE<br>MSD | Various Methods |
| [212] | ABIDE 1 | rs-fMRI | AAL Atlases + DPARSF Pipeline | VAE | VAE with 3 Layers | Adadelta | Proposed Loss Function | -- |
| [213] | ABIDE 1 | rs-fMRI | CC200 Atlases + CCS Pipeline | LSTM | LSTM (1) + Pooling (1) + FC (1) | Adadelta | BCE | Sigmoid |
| [214] | ABIDE | rs-fMRI<br>Phenotypic Info | CC200 Atlases | SAE | SAE (3) + Sigmoid | Proposed Opt. L-BFGS | MSE | Clustering |
| [215] | ABIDE | rs-fMRI | -- | SAE | LSTM (1) + Pooling (1) + FC (1) | Adadelta | BCE | Sigmoid |
| [216] | ABIDE 1 | rs-fMRI | AAL Atlases + CCS Pipeline | LSTM | LSTM (2) + Pooling (1) + FC (2) | Adam | BCE<br>MSE | Sigmoid |

| Ref | Dataset | Modality | Atlas/Pipeline | Model | Architecture | Optimizer | Loss | Classifier |
|---|---|---|---|---|---|---|---|---|
| [217] | ABIDE 1 | rs-fMRI | Different Atlas | DANN | 3 MLP (1 Dropout + 4 Dense) + Self-Attention (3) + Fusion (3) + Aggregation + Dense (1) + ReLU, ELU, and Tanh | -- | SE | Sigmoid |
| [218] | ABIDE | rs-fMRI | AAL Atlas + CCS Pipeline | SSAE | 3 SSAE Layers | Gradient Descent | Proposed Loss Function | Softmax Regression |
| [219] | ABIDE 1 +II | rs-fMRI | Different | 1D-CNN | CNN (1) + Pooling (1) + FC (1) | Adam | -- | Softmax |
| [220] | ABIDE 1 | rs-fMRI | Different Pipelines | Various Models | CNN (6) + Pooling (4) + BN (2) + FC (2) | Adam | Propose Loss Function | Sigmoid |
| [221] | ABIDE-II | rs-fMRI | NA | 1D-CAE | Encoder (4) + Decoder (4) + CNN (2) + pooling (2) + FC (2) | -- | -- | -- |
| [222] | ABIDE | rs-fMRI | CCS Pipeline | AlexNet | Standard Architecture | -- | CE | Softmax |
| [223] | ABIDE | rs-fMRI | Different Pipelines | ASDDiagNet | Proposed DiagNet | -- | -- | SLP |
| [224] | ABIDE 1 | rs-fMRI | SCSC Atlas + CPAC Pipeline | Auto-ASD | Proposed Auto-ASD-Network | -- | NLLF | SVM |
| [225] | ABIDE 1 | rs-fMRI | CC400 Atlas + CPAC Pipeline | 2D-CNN | CNN (7) + Pooling (7) + FC (3) | -- | -- | MLP |
| [226] | ABIDE | rs-fMRI | NA | CNN-AE | Proposed SDAE-CNN with 7 Layes CNN | -- | -- | Softmax |
| [227] | ABIDE 1 | rs-fMRI | NA | 3D-FCNN | CNN (9) + PReLU + FC (3) | SGD | CE | Softmax |
| [228] | ABIDE 1 | rs-fMRI | AAL Atlas + DPARSF Pipeline | SSAE | 2 Layers SSAE | -- | -- | Softmax |
| [229] | ABIDE 1 | rs-fMRI | NA | 3D-CNN | CNN (7) + Pooling (3) + FC (2) + Log-Likelihood Activation | SGD | MNLL | -- |
| [230] | ABIDE 1 | rs-fMRI | HO Atlas + CPAC Pipeline | GCN | GCN with ReLU and Sigmoid | -- | CE | Softmax |
| | | Phenotypic Info | | AE | SAE wth Tanh Activation | | MSE | |
| [231] | ABIDE 1 | rs-fMRI | CC200 Atlas + CCS Pipeline | LSTM | Proposed Method | Adadelta | BCE | Sigmoid |
| | | Phenotypic Info | | | | | MSE | |
| [232] | ABIDE 1 | rs-fMRI | CC200 Atlas + CPAC Pipeline | SDAE | Proposed 2-SDAE-MLP Network | -- | MSE | Softmax |
| | | s-MRI | | | | | | |
| | | Phenotypic Info | | | | | | |
| [233] | Clinical | rs-fMRI | NA Atlas | 3D-CNN | CNN (2) + ReLU + Pooling (2) + FC (2) | SGD | BCE | Sigmoid |
| | | Fetal BOLD fMRI | | | | | | |
| [234] | ABIDE 1 | rs-fMRI | AAL Atlas | DBN | DBN with 5 Hidden Layers | -- | -- | LR |
| | ABIDE-II | s-MRI | | | | | | |
| [235] | IMPAC | rs-fMRI | Different Atlases | Various Models | Dense (5) + LReLU | Adam | BCE | Various Methods |
| | | s-MRI | | | | | | |
| [236] | ABIDE 1 | rs-fMRI | AAL, CC200, Destrieux Atlas + CPAC Pipeline | SAEs | 5 [ AE (3) + MLP (2)] + Softmax | -- | -- | Softmax |
| | | s-MRI | | | | | | |
| [237] | NDAR | rs-fMRI | NA | CNN | CNN (5) + ReLU + Pooling (2) + FC (5) | -- | CE | Softmax |
| | | s-MRI | | | | | | |
| [238] | NDAR | All Modalities | Implement the Proposed Atlas | SAE | 34 [ SAE network (2)] | PSVM | L-BFGS | Probabilistic SVM (PSVM) |
| [239] | NDAR | s-MRI | NA | DDUNET | Proposed DDUNET with 11 blocks and ReLU | SGD | CE | -- |
| [240] | ABIDE 1 | s-MRI | NA | SNCAE | Proposed SNCAE Newtork | -- | -- | Softmax |

| Ref | Dataset | Modality | Atlas/Pipeline | Model | Architecture | Optimizer | Loss | Classifier |
|---|---|---|---|---|---|---|---|---|
| | NDAR/Pitt | | | | | | | |
| | NDAR/IBIS | | | | | | | |
| [241] | ABIDE 1 | s-MRI | Destrieux | SpAE | SpAE with 2 Networks | -- | MSE | Softmax |
| [242] | HCP ABIDE 1 | s-MRI | Desikan–Killia (DDK) Atlas | DEA | AE (3) + SELU | Adam | Sum of MSE + 2 CE + CC | -- |
| [243] | ABIDE CombiRx | s-MRI | NA | DCNN | CNN (6) + ReLU + Pooling (6) + FC (4) | Adam | BCE | Sigmoid |
| [244] | ABIDE-II | s-MRI | DKT Atlas | CNN | Proposed FastSurfer CNN Network | Adam | Logistic & Dice Losses | Softmax |
| [245] | ABIDE 1 | s-MRI | Different | 3D-CNN | CNN (3) + ReLU + Pooling (3) + FC (2) | Adadelta | CE | Softmax |
| [246] | Clinical | s-MRI | NA | UNet | DCNN (7) + ReLU + Pooling (2) + BN (6) | SGD | weighted CE | Softmax |
| [247] | ABIDE 1 | rs-fMRI | BN Atlas + CPAC Pipeline | CNN-RNN | CNN (4) + GRU (2) + ReLU + Pooling (2) + FC (5) | Adam | BCE | Sigmoid |
| [248] | Clinical | fNIRS | NA | CNN-LSTM | Proposed 1D-CNN LSTM with ReLU Activation | Adam | CCE | Bagging |
| [249] | Clinical | fNIRS | NA | CGRNN | CNN (3) + ReLU + Pooling (1) + GRU (1) + FC (1) | Adam | BCE | Sigmoid |
| [250] | Different Datasets | s-MRI | Different Atlas | CNN | Variation of the U-net Convolutional Architecture | Adam | Proposed Loss Function | -- |
| [251] | ABIDE 1+II | rs-fMRI | HO Atlas + CPAC Pipeline | 3D-CNN | CNN (2) + ELU + Pooling (2) + FC (2) | SGD | BCE | MV |
| [252] | ABIDE 1 | rs-fMRI Phenotypic Info | CPAC Pipeline | CNN-RNN | CNN (8) + Conv-BiLSTM (2) + Sigmoid + Pooling (1) + FC (1) | Adam | CE | Softmax |
| [253] | ABIDE 1 | rs-fMRI s-MRI | NA | AE | Proposed AE with 7 Layers | -- | -- | DNN |
| [254] | ABIDE 1 | rs-fMRI | CC200 Atlas + CPAC Pipeline | CapsNet | Standard Architecture | Adam | Proposed Loss Function | K-Means Clustering |
| [255] | ABIDE | rs-fMRI | CCS Pipeline | convGRU-CNN3D | convGRU+ 3D CNN | Adam | CE | -- |
| [256] | ABIDE | rs-fMRI | CC200 Atlas + CPAC Atlas | 1D-CNN | Conv (3) + Pooling (3) + FC (1) | -- | -- | Softmax |
| [257] | ABIDE | rs-fMRI | CC200, AAL, Dosenbach Atlas + CPAC Pipeline | SDA | 3 DAEs (Each 3 Hidden Layers) | -- | -- | Different Classifier |
| [258] | ABIDE 1 | rs-fMRI | AAL Atlas + CPAC Pipeline | Proposed CNN Method | Conv (1) + Max (1) + Res-blocks (4) + Average Pooling (1) + 1 FC + ReLU | Mini-Batch SGD | Softmax CE | 4 FC |
| [259] | ABIDE | MRI | NA | DNN | Different Configurations | -- | -- | -- |
| [260] | ABIDE | rs-fMRI | DPARSF Pipeline | RBM | Proposed Architecture | -- | -- | SVM |
| [261] | ABIDE 1 ABIDE II | rs-fMRI | GCA Atlas HO Atlas DCA Atlas | 1D-CNN | Conv (3) + Pooling (1) + FC (2) + ReLU Conv (3) + BN (3) + Pooling (4) + FC (1) + ReLU | -- | CE | Softmax |
| [262] | IMPAC | sMRI rs-fMRI | Proposed Atlas | 2D-CNN | Conv (3) + Pooling (1) + FC (2) + ReLU Conv (3) + BN (3) + Pooling (4) + FC (1) + ReLU | Proposed Opt | -- | Softmax |
| [263] | ABIDE 1 | MRI | HO, SSA Atlas | 3D-CNN | Conv (3) + Pooling (3) + FC (2) + ReLU | Adadelta | CE | Softmax |
| [264] | ABIDE 1 | rs-fMRI | CC400 Atlas + DPARSF Pipeline | GCN | Proposed Architecture | -- | -- | Majority Voting |
| [265] | ABIDE | fMRI | CC200 Atlas | GAN | Encoder, Generator, Discriminator Networks | Adam | -- | 3-Layer DNN |
| [266] | ABIDE | rs-fMRI | AAL Atlas | PreTrained Artcitectures | Densenet201, GoogleNet, Resnet101, Resnet18 | -- | -- | SVM, KNN |
| [267] | ABIDE 1 | sMRI | NA | ResNet50 | Standard Architecture | Adam | -- | Sigmoid |
| [268] | ABIDE II | rs-fMRI | AAL Atlas | Proposed 3D CNN Method | 3D Convolutional Autoencoders | -- | -- | SVM |

| Ref | Dataset | Modality | Atlas/Pipeline | Model | Architecture | Optimizer | Loss | Output |
|---|---|---|---|---|---|---|---|---|
| [269] | ABIDE 1 | rs-fMRI, MRI | Various Atlas | CNN-AE | Hidden Layers (4) | Adam | CE | Softmax |
| [270] | ABIDE 1 / ABIDE II | sMRI | DK Atlas | DenseNet | 4 Dense Blocks | -- | -- | LDA |
| [271] | ABIDE 1 | fMRI, sMRI | NA | 2D-CNN and 3D-CNN | Conv (3) + Pooling (3) + FC (2) | -- | -- | Different Methods |
| [272] | NDAR | Task fMRI | HO Atlas | 2D CNN | Conv (3) + Pooling (3) + FC (2) | SGD | -- | Softmax |
| [273] | ABIDE | rs-fMRI | CC200 Atlas + CCS Pipeline | cGCN | Conv (5) + RNN (Temporal Average Pooling Layer) | Adam | -- | Softmax |
| [274] | ABIDE | rs-fMRI | CC200 Atlas + CPAC Pipeline | 1D-CNN | Conv (3) + Pooling (3) + FC (1) | -- | -- | Softmax |
| [275] | ABIDE 1 | rs-fMRI | CC400 Atlas + CPAC Pipeline | DNN | Hidden Layers (2) | Adam | BCE | Softmax |
| [276] | ABIDE 1 | rs-fMRI | AAL Atlas + CPAC Pipeline | FCNN | FC (2) + BN (2) + LeakyReLu | Adam | BCE | Sigmoid |
| [277] | ABIDE 1 | rs-fMRI | CC200 Atlas + CPAC Pipeline | DCAE | Encoder: 4 CNN Networks [Conv (4) + BN (3) + Pooling (2) + ReLU], Decoder: 4 CNN Networks [Reverse Configuration] | -- | -- | FC Layer + Dense Layer |
| [278] | ABIDE 1 | rs-fMRI | CC400 Atlas + CPAC Pipeline | 2D-CNN | Conv (2) + EvoNorm-S0 (2) + Tanh + FC Layer (1) | Adam | BCE | Sigmoid |
| [279] | ABIDE 1 | sMRI | SRI24 Atlas | ResNet | Conv Blocks (5) | Adam | BCE | FC Network |
| [280] | ABIDE 1 | rs-fMRI | CPAC Pipeline | CNN-AE | AE Network + 1D-Conv (2) + BN (2) + ReLU + Pooling (2) + FC (2) | | | Sigmoid |
| [281] | ABIDE | fMRI | AAL Atlas | FC | Pooling (2) + Conv (3) + FC (1) + ReLU | Adam | -- | Softmax |
| | | | | ALFF Net | Pooling (2) + 3D-Conv (4) + FC (1) + ReLU | | | |
| [282] | ABIDE 1 | ??? | DK Atlas | CGTS-GAN | Proposed Architecture | -- | -- | -- |
| [283] | ABIDE 1 / ABIDE II | Rs-fMRI | AAL Atlas + DPARSF Pipeline | CNN | NA | -- | -- | Softmax |
| [284] | ABIDE 1 | rs-fMRI | CPAC Pipeline | SAE | Two unsupervised Sparse AEs | -- | -- | Sypervised AE |
| [285] | ABIDE 1 | rs-fMRI | BASC Pipeline | CNN-RNN | Different Architectures | -- | -- | Sigmoid |
| [286] | ABIDE 1 | rs-fMRI | MSDL Pipeline | 1D-CNN | Conv (4) + BN (3) + Pooling (3) + FC (2) + ReLU | -- | -- | Sigmoid |
| [287] | ABIDE 1 | rs-fMRI | CCS Pipeline | Inception-ResNetV2 | InceptionResNetV2 Architecture + Pooling Layers | SGD | -- | Softmax |
| [288] | ABIDE 1 / ABIDE II | rs-fMRI | NA | ResNet-50 | Standard Architecture | -- | -- | Softmax |
| [289] | ABIDE 1 | rs-fMRI | CC200 + CPAC Pipeline | ASD-SAENet | SAE + DNN | Adam | CE | Softmax |
| [290] | ABIDE 1 | rs-fMRI | BASC, CC200, and AAL Atlas + CPAC Pipeline | DNN | Hidden (2) + ReLU | -- | -- | Sigmoid |
| [291] | ABIDE 1 | rs-fMRI | CC200 Atlas + CPAC Pipeline | DBN | 3 RBM Layers | RMSProp | CE | Softmax |
| [292] | Different Dataset | sMRI and fMRI | AAL Atlas | 2D-CNN | Conv (1) + FC (3) + BN (3) + ReLU | Adam | -- | Softmax |
| [293] | ABIDE 1 | rs-fMRI | AAL Atlas + DPARSF Pipeline | CNNGLasso | Conv Layer + FC Layer | Adam | Proposed Loss Function | Softmax |
| [294] | ABIDE 1 / ABIDE II | sMRI, fMRI | Various Atlas | GCN | p-GCN + s-GCN + ss-GCN Networks | -- | -- | -- |
| [295] | KKI | rs-fMRI, DTI | AAL Atlas | MGCN | Proposed Architecture | SGD | | ANN |

| Ref | Dataset | Modality | Atlas | Method | Architecture | Optimizer | Loss | Output |
|---|---|---|---|---|---|---|---|---|
| | HCP | | | | | | Combined Loss Function | |
| [296] | Clinical | rs-fMRI, DTI | AAL Atlas | LSTM-ANN | LSTM (2) + Both the P-ANN and the A-ANN Have 2 Hidden Layers + ReLU | Adam | Proposed Loss Function | Attention-Weighted Average |
| [328] | NDAR | T-fMRI, sMRI | HO | 1D-CNN | CNN (2) + ReLU + Pooling (2) + FC (1) | SGD | -- | Softmax |
| [329] | ABIDE | rs-fMRI | CCS | 3D-CNN | CNN (3) + ReLU + Pooling (3) + FC (1) | Adam | -- | Sigmoid |
| [330] | ABIDE | rs-fMRI | 6 Atlases + CPAC | Multi-Atlas Graph Convolutional Network Method (MAGCN) | GCN Model | -- | Cross Entropy | Stacking Ensemble Learning Method + Ridge Classifier |
| [331] | ABIDE I | rs-fMRI | -- | MHATC | Multi-Head Attention Encoder (MHAE) + Temporal Consolidation Module (TCM) | -- | Cross Entropy | Pooling (1) + FC (1) |
| [332] | ABIDE I | rs-fMRI | CC400 + CPAC | Simplified VAE Unsupervised Pretraining And MLP Supervised Fine-Tuning | Hidden Layers (3) | RMSProp | Cross Entropy | Softmax |
| [333] | NDAR | sMRI, fMRI | Talairach | Deep Fusion Classification Network (DFCN) | 2 Stacked Autoencoder With Non-Negativity Constraint (SNCAE) | -- | -- | Softmax |
| [334] | ABIDE | sMRI | -- | 2D CAM, 3D CAM, 3D Grad-CAM | Proposed Architectures | -- | Cross Entropy | Classification Output Layer |
| [335] | ABIDE | rs-fMRI | HO + CPAC | Invertible Dynamic GCN (ID-GCN) | Three Invertible Blocks (2 Different GCN) | -- | Cross Entropy | Softmax |
| [336] | ABIDE | rs-fMRI | AAL, HO, MODL | Functional Graph Discriminative Network (FGDN) | Functional Graph Construction Layer (1) + Graph Conv Layers (2) + PReLU + FC Layer (1) | Adam | Proposed Loss Function | Sigmoid |
| [337] | ABIDE | rs-fMRI, Phenotypic | CPAC + HO | Combined DFS and GCN Method | Sparse One-To-One Linear Layer + Hidden Layers (3) + Graph Conv Layers + ReLU + Dropout Layer | Adam | Proposed Loss Function | Softmax |
| [338] | ABIDE | rs-fMRI, Phenotypic | -- | Adaptive Multi-Layer Aggregation Graph Convolutional Network (AMA-GCN) | Proposed Architectures | Adam | Fusion Loss | Softmax |
| [339] | ABIDE I | rs-fMRI, Phenotypic | CPAC + HO | DeepGCN | 16 Layers GCN + Dropout + ResNet Units + DropEdge Strategy | -- | -- | Softmax |
| [340] | ABIDE | rs-fMRI | AAL, CC200 + DPARSF | Multi-Scale Graph Representation Learning (MGRL) Framework | Multi-Scale FCNs Construction + FCNs Representation Learning Via Multi-Scale GCNs + Multi-Scale Feature Fusion and Classification | Adam | Cross Entropy | Softmax |
| [341] | ABIDE | rs-fMRI | -- | CNN | CNN (3) + ReLU + Pooling (2) + FC (2) | -- | -- | Softmax |
| [342] | ABIDE I | rs-fMRI | -- | SSAE + MLP | Dense Layers (4) + ReLU + Dense Layers (4) | -- | Cross Entropy | Softmax |
| [343] | ABIDE | rs-fMRI | AAL, CC200 + CPAC | Multi-View Graph Convolutional Neural Network (MVS-GCN) | Graph Structure Learning (GSL) + Multi-Task Graph Embedding Learning for Different Views of Brain Networks (MVL) + View Consistency Regularization (VCR) and the Prior Subnetwork Structure Regularization (SNR). | Adam | Proposed Loss Function | -- |